\documentclass[sigconf,nonacm]{acmart}

\usepackage{amsmath,booktabs,multirow,makecell,pifont}
\usepackage{graphicx}
\usepackage{xcolor}
\usepackage{xspace}
\usepackage{marvosym}
\usepackage{algorithm}
\usepackage[noend]{algpseudocode}

\newcommand{\scalarlens}{\textsc{ScalarLens}\xspace}
\newcommand{\num}{\mathrm{num}}
\newcommand{\catg}{\mathrm{cat}}

\newcommand{\singlecoltablesize}{\scriptsize}
\newcommand{\singlecoltableformat}{\singlecoltablesize\setlength{\tabcolsep}{2.4pt}\renewcommand{\arraystretch}{1.08}}
\newcolumntype{P}[1]{>{\centering\arraybackslash}p{#1}}
\definecolor{stdblue}{RGB}{70,104,128}
\newcommand{\aucstd}[2]{\ensuremath{#1_{\textcolor{stdblue}{\scriptscriptstyle\pm #2}}}}
\makeatletter
\renewcommand{\country}[1]{\global\@ACM@countrypresenttrue}
\makeatother

\begin{document}

\title{ScalarLens: Numerical Embeddings with Stable Coordinates and Contextual Responses for CTR Prediction}

\author{Heng Yao\textsuperscript{\Letter}}
\affiliation{%
  \institution{Ant Group}
  \country{}
}

\author{Tianying Liu}
\affiliation{%
  \institution{Independent Researcher}
  \country{}
}

\author{Yulou Shu}
\affiliation{%
  \institution{Alibaba Inc.}
  \country{}
}

\author{\makebox[0pt][c]{Yong He, Chuan Yuan, Kaibin Qiu}\\
  \makebox[0pt][c]{Guowei Chen, Jiayu Zhao}}
\affiliation{%
  \institution{Ant Group}
  \country{}
}

\author{Siyun Hou\textsuperscript{\Letter}}
\affiliation{%
  \institution{Henan Polytechnic University}
  \country{}
}

\renewcommand{\shortauthors}{Yao et al.}

\begin{abstract}
Numerical embeddings for click-through rate (CTR) prediction are built on a convenient but restrictive premise: a scalar has one representation. This premise conflates where a value lies with what it means for the current sample. On the Criteo validation split, the same numerical interval carries residual click evidence with opposite signs across categorical and numerical contexts, even after additive main effects are removed. Production pipelines compound this mismatch because externally normalized features require transformations and statistics to remain synchronized between training and serving.

We introduce \scalarlens, a numerical embedding that preserves \emph{what a value is} while adapting \emph{how it should be interpreted}. A monotone local mesh constructs a stable coordinate from the focal scalar alone; bounded low-rank dynamics then produce a contextual response without moving that coordinate or replacing categorical tokens and the CTR backbone. In a 1,539-run primary evaluation covering 19 representations, three datasets, nine backbones, and three seeds, \scalarlens ranks first in 25 of 27 settings on original numerical scales and second in the remaining two. Matched ablations show that scale correction, additional local capacity, and generic conditioning do not reproduce the gain. A controlled study further recovers categorical, numerical, and mixed response mechanisms under context shift while the focal coordinate remains exactly invariant. A complete rerun under shared standardization retains significant advantages over DEER, DAES, and NaryDis, showing that the result is not explained by tolerance to raw scales alone. \scalarlens therefore recasts numerical embedding as a measurement problem: coordinates belong to values, while predictive responses belong to values in context.
\end{abstract}



\maketitle

\begingroup
\makeatletter
\def\@makefnmark{\hbox{\@textsuperscript{\Letter}}}
\footnotetext{Corresponding authors: Heng Yao and Siyun Hou. E-mail: \href{mailto:yaoheng.cs@qq.com}{yaoheng.cs@qq.com} / \href{mailto:housiyun@hpu.edu.cn}{housiyun@hpu.edu.cn}.}
\makeatother
\endgroup

\section{Introduction}

Click-through rate (CTR) prediction underpins sponsored search, recommendation, and content ranking. Numerical representation in these systems is usually treated as a problem of geometry in one dimension: for field $i$ and value $x_i$, construct
\begin{equation}
e_i=f_i(x_i).
\label{eq:static_encoder_intro}
\end{equation}
Existing methods impose discretized, local, spectral, hierarchical, or distributional geometries~\cite{guo2021autodis,gorishniy2022numemb,cheng2022deer,chen2022narydis,shen2023dae}. They differ in how they locate a value, yet Equation~\eqref{eq:static_encoder_intro} assigns the same vector before feature interaction in every sample. It therefore treats coordinate and predictive meaning as one object.

There is also an operational reason to revisit this interface. Production numerical fields are commonly generated by different upstream jobs, refreshed at different cadences, and revised under independent feature versions. An external normalizer therefore adds more than two fitted moments: its transformation, lineage, and version must remain synchronized across offline training, backfills, and online serving. Materializing normalized columns consumes storage and I/O, whereas computing them on demand repeats preprocessing for every example and adds another consistency requirement between training and serving. We consequently treat direct learning from values in their original scale as a deployment requirement, not merely a stress test. This choice does not eliminate schema checks, clipping, or drift monitoring. It removes the need to manage a separate normalized representation of each feature.

\begin{figure}[t]
  \centering
  \includegraphics[width=\linewidth]{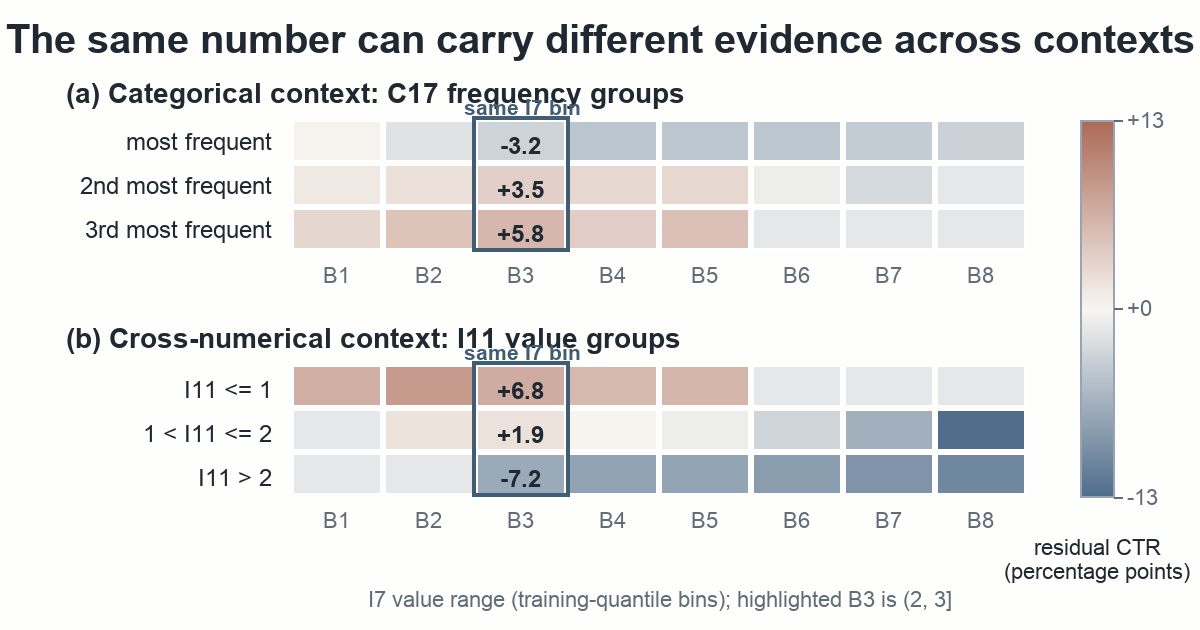}
  \caption{\textbf{The same numerical coordinate can carry different predictive evidence across contexts.} Both panels use the Criteo validation split and target field $I7$. Each cell shows observed CTR minus an empirical Bayes additive estimate of the main effects of the target bin and context group. Panel (a) groups rows by categorical field $C17$, and panel (b) groups them by ranges of numerical field $I11$. The highlighted interval $(2,3]$ reverses sign under both sources. Pair selection and group boundaries use training data only, and cells with fewer than 500 validation examples are omitted. The panels establish the representation problem but do not imply that both context sources have equal downstream utility. No CTR backbone, test label, or \scalarlens output is used.}
  \Description{Two heatmaps show residual click-through rate across eight ranges of Criteo numerical field I7. The first groups rows by categorical field C17 and the second by ranges of numerical field I11. In the highlighted I7 interval from two to three, residuals change sign across rows in both panels.}
  \label{fig:motivation}
\end{figure}

Figure~\ref{fig:motivation} examines this assumption without using a CTR model. An empirical Bayes additive estimator first removes the main effects of the target bin and context group. For the same Criteo $I7$ interval $(2,3]$, the remaining CTR associations are $-3.2$, $+3.5$, and $+5.8$ percentage points across groups of $C17$, and $+6.8$, $+1.9$, and $-7.2$ across ranges of $I11$. The numerical value and its order do not change, but the associated evidence reverses sign. This observation motivates a representation that preserves the scalar coordinate while adapting its response to the sample.

The mismatch arises at the representation interface. A downstream factorization, cross, or attention module may learn interactions, but an encoder invariant to context has already committed to one numerical token. When $p(y\mid x_i,c)$ varies with context $c$, each consumer must reconstruct the missing conditional structure. We instead require
\begin{equation}
e_i^{\num}=\Psi_i\!\left(b_i(x_i),c\right),
\label{eq:context_factorization_intro}
\end{equation}
where $b_i$ is a \emph{stable coordinate} determined only by the scalar and $\Psi_i$ is a separate response conditioned on the sample. Stable means invariant to the context in the same sample, not fixed during learning. Thus coordinates belong to values, whereas responses belong to pairs of values and contexts. Context cannot alter coordinate construction, and neither path replaces categorical tokens or the CTR backbone.

Contextual refinement and conditional numerical encoding already exist~\cite{wang2021contextnet,wang2022frnet,liu2026daes}; our claim is not the first use of context. The unresolved issue is \emph{where context acts}. Generic refinement can rewrite a token, while conditional interpolation can change a representation derived from the coordinate. \scalarlens defines this boundary explicitly: a monotone mesh inspired by DEER anchors the scalar, conditions that preserve field identity drive a bounded low-rank response, and a readout that returns only numerical embeddings retains both parts. The construction follows model reduction in computational physics, in which the scalar specifies an observation point and the sample supplies forcing conditions.

Our contributions are:
\begin{itemize}
  \item \textbf{Problem.} Validation data reveal opposing evidence for identical scalar intervals across contexts. At the same time, heterogeneous feature pipelines make separately materialized or served normalization an avoidable operational dependency.
  \item \textbf{Method.} \scalarlens consumes values in their original scale and separates a monotone coordinate determined only by the scalar from a bounded contextual response that preserves field identity. Its module interface changes only the numerical embeddings.
  \item \textbf{Evidence.} A complete matrix of 1,539 runs places \scalarlens first in 25 of 27 combinations of datasets and backbones. Ablations, controlled recovery, rarity tests, and a secondary audit under shared standardization isolate the mechanism and rule out tolerance to input scale as the sole explanation.
\end{itemize}

\section{Related Work}

Prior work follows three lines: numerical representation, general contextual refinement, and conditional numerical encoding. Table~\ref{tab:innovation} distinguishes where context acts rather than treating every design choice as an advantage.

\subsection{Numerical Representation for CTR}

AutoDis learns soft assignments over meta-embeddings~\cite{guo2021autodis}; PLE preserves position within each bin; Periodic and PLR expose frequency components~\cite{gorishniy2022numemb}; and PLE-B adds a residual shortcut~\cite{gorishniy2025tabm}. B-splines provide smooth local support~\cite{shtoff2024basis}. DEER interpolates adjacent nodes on a learned mesh~\cite{cheng2022deer}; NaryDis represents several numeral system scales~\cite{chen2022narydis}; and DAE uses distances to distributional anchors~\cite{shen2023dae}. Recent studies revisit these structures, uncertainty, and semantic tokens for features and values~\cite{gao2025revisiting,kartashev2026uncertainty,koloski2025llmemb}. Despite different bases, each scalar receives the same token in every sample.

These bases encode locality, resolution, density, or smoothness, but even a precise univariate location remains invariant to the response required by each sample.

Consumer choice can confound representation quality because WideDeep, DeepFM, NFM, PNN, FiBiNET, DCN/DCNv2, and AutoInt impose different linear, product, cross, and attention biases~\cite{cheng2016widedeep,guo2017deepfm,he2017nfm,qu2016pnn,huang2019fibinet,wang2017dcn,wang2021dcnv2,song2019autoint}. Protocol studies warn against conclusions from one consumer~\cite{zhu2021bars,rubachev2025tabred,grinsztajn2022trees,mcelfresh2023neural}; we therefore vary nine backbones while changing only the numerical token. Pretrained tabular models address the distinct goal of transfer across tables~\cite{hollmann2025tabpfn,qu2025tabicl,ma2025tabdpt,zhang2025limix,qu2026tabiclv2,wang2026limix2m,grinsztajn2026tabpfn3,hosseinzadeh2026tabdptturbo}.

\subsection{Contextual Feature Representation}

Contextualization usually targets general feature refinement. FiLM predicts featurewise affine transformations from conditioning information~\cite{perez2018film}; ContextNet and FRNet adapt embeddings through contextual layers or dimension level gates~\cite{wang2021contextnet,wang2022frnet}. Because they act on generic latent features, these methods may rewrite categorical and numerical fields together. They strengthen interaction modeling, but do not preserve an explicit scalar coordinate whose order can be examined independently.

This boundary matters for numerical embedding: gains from refining every field may reflect categorical recalibration rather than better representations of numbers. Our matched ContextFiLM ablation therefore lets selected categorical fields condition only numerical outputs while fixing output shape and the consumer. \scalarlens instead retains the ordered coordinate and restricts context to a separate bounded response.

\subsection{Conditional Numerical Encoding}
\label{sec:conditional-representations}

DAES~\cite{liu2026daes} is the closest conditional encoder. Selected categorical fields modulate interpolation weights over a maintained quantile coordinate for non-IID streams. \scalarlens instead learns a coordinate determined only by the focal scalar and applies categorical and auxiliary numerical evidence to a distinct bounded response. It accepts original values without external normalization or maintained streaming quantiles. The distinction is what context may change: interpolation weights in DAES, versus a response downstream of an invariant coordinate in \scalarlens.

\paragraph{Innovation positioning.}
Table~\ref{tab:innovation} separates unconditioned scalar encoders, DAES's conditional quantile interpolation, and all field refiners. \scalarlens instead fixes the coordinate response boundary, returns numerical tokens, and needs neither streaming statistics nor a specialized consumer.

\begin{table}[t]
\caption{Positioning of numerical encoders and contextual refiners by their treatment of coordinates and responses. Arrows denote context source $\rightarrow$ adaptation location; entries describe mechanisms rather than capabilities.}
\label{tab:innovation}
\centering
\singlecoltablesize
\setlength{\tabcolsep}{0.8pt}
\renewcommand{\arraystretch}{1.10}
\resizebox{\columnwidth}{!}{%
\begin{tabular}{lcccc}
\toprule
Method & Scalar coordinate & Context action & Statistics & Output \\
\midrule
AutoDis~\cite{guo2021autodis} & Soft bins & None & None & Numerical \\
PLE~\cite{gorishniy2022numemb} & Linear bins & None & Bin edges & Numerical \\
Periodic / PLR~\cite{gorishniy2022numemb} & Periodic & None & None & Numerical \\
B-spline~\cite{shtoff2024basis} & Splines & None & Knots & Numerical \\
DEER~\cite{cheng2022deer} & Learned mesh & None & None & Numerical \\
NaryDis~\cite{chen2022narydis} & Multi-radix & None & None & Numerical \\
DAE~\cite{shen2023dae} & Anchor distance & None & Anchors & Numerical \\
DAES~\cite{liu2026daes} & Quantiles & Sel. cat. $\rightarrow$ weights & Online quantiles & Numerical \\
ContextNet / FRNet~\cite{wang2021contextnet,wang2022frnet} & Generic features & All $\rightarrow$ features & None & All fields \\
\midrule
\textbf{\scalarlens (ours)} & Local mesh & Cat. + num. $\rightarrow$ response & None & Numerical \\
\bottomrule
\end{tabular}}
\end{table}

Experiments separately test the module interface, generic conditioning, coordinate invariance, and response adaptation.

\section{\scalarlens: Stable Coordinates and Contextual Responses}
\label{sec:method}

\subsection{One Numerical Embedding, Two Coupled Mechanisms}

\scalarlens decomposes the numerical interface into two inspectable objects:
\begin{equation}
e_i^{\num}=\Psi_i\!\left(e_i^{\mathrm{loc}}(x_i),
\mathcal{C}(\mathbf{x}^{\num},\mathbf{E}^{\catg})\right),
\label{eq:contextual_factorization}
\end{equation}
where $e_i^{\mathrm{loc}}$ depends only on the focal scalar and remains an explicit readout input. Context affects a separate response state that adapts how the location is presented to the task. Their constrained composition can be tested directly through ablation.

The design draws on model reduction~\cite{quarteroni2016reduced}: a scalar locates a point on a mesh, surrounding fields provide forcing conditions, and a compact state approximates the response through a fixed number of updates. This analogy supplies an inductive bias; it does not claim that CTR data obey a physical law.

The analogy yields three constraints. The observation point is computed before forcing, so context cannot relocate it. A shared low-rank operator communicates across fields without a second interaction backbone. Bounded updates of fixed depth replace an unconstrained iterative solver. Ablations test each mechanism.

For $N$ numerical fields and $C$ categorical fields, lookup tables produce $\mathbf{E}^{\catg}\in\mathbb{R}^{C\times d}$. \scalarlens learns
\begin{equation}
\Phi:(\mathbf{x}^{\num},\mathbf{E}^{\catg})\mapsto
\mathbf{E}^{\num}\in\mathbb{R}^{N\times d},
\label{eq:module_boundary}
\end{equation}
and the unchanged backbone consumes $[\mathbf{E}^{\num};\mathbf{E}^{\catg}]$. The encoder reads the sample context but returns only numerical tokens; categorical embeddings bypass it unchanged.

\begin{figure}[t]
  \centering
  \includegraphics[width=\columnwidth]{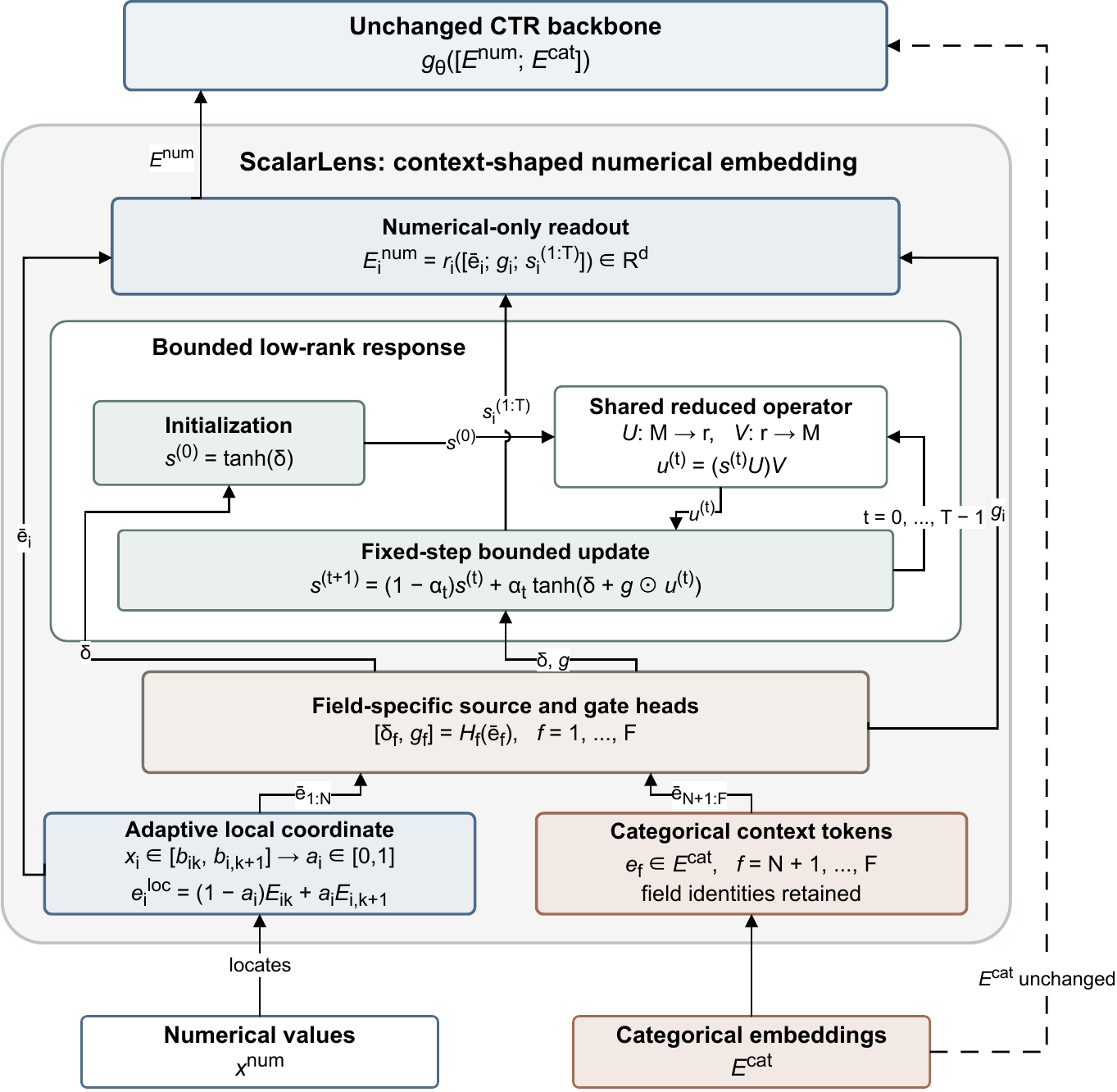}
  \caption{\textbf{\scalarlens couples stable coordinates with contextual responses.} A numerical value alone selects an adaptive local coordinate. Field tokens normalized by their root mean square produce identifiable drives $\delta_f$ and gates $g_f$ for a bounded response with $T$ updates and a shared low-rank operator $UV$. The readout retains the focal coordinate $\bar e_i$ together with $g_i$ and $s_i^{(1:T)}$, and returns only numerical embeddings. Original categorical embeddings bypass the module and enter the unchanged CTR backbone.}
  \Description{A computation graph read from bottom to top. Numerical values enter an adaptive local coordinate block, and categorical embeddings enter a context block. Separate source and gate heads for each field feed a bounded low-rank response. The readout emits only numerical embeddings, while original categorical embeddings bypass the module and enter an unchanged CTR backbone.}
  \label{fig:overview}
\end{figure}

\subsection{Stable Numerical Coordinates}

For numerical field $i$, let $[l_i,u_i]$ be its training range and $K$ the number of intervals. Width logits are converted into positive normalized widths and cumulative boundaries:
\begin{align}
p_{ik}&=\operatorname{softmax}(\mathbf{w}_i/\tau)_k,
&q_{ik}&=\frac{p_{ik}+\epsilon}{1+K\epsilon},\nonumber\\
b_{i0}&=l_i,
&b_{ik}&=l_i+(u_i-l_i)\sum_{j=1}^{k}q_{ij}.
\label{eq:monotone_mesh}
\end{align}
Hence $b_{i0}<\cdots<b_{iK}$. For $x_i\in[b_{ik},b_{i,k+1}]$, clipping outside the range,
\begin{equation}
e_i^{\mathrm{loc}}=(1-a_i)E_{ik}+a_iE_{i,k+1},\qquad
a_i=\frac{x_i-b_{ik}}{b_{i,k+1}-b_{ik}}.
\label{eq:local_interpolation}
\end{equation}
The interpolation, inspired by DEER~\cite{cheng2022deer}, is continuous, ordered, and local. Context cannot participate in interval selection or interpolation.

\paragraph{Input contract for original numerical scales.}
The encoder receives values in their serving units. Training ranges $[l_i,u_i]$ are stored with the checkpoint, and inference clips to their endpoints. The pipeline therefore need not materialize normalized copies or reproduce a mean and variance transformation online. Source validation and drift monitoring remain necessary; the benefit is a shorter feature lifecycle, not the elimination of data governance.

\paragraph{Coordinate regularity.}
Equal focal values produce identical coordinates in every context. Adjacent intervals share a node, the positive floor prevents zero width, and clipping defines endpoints outside the training range. Context is introduced only after this coordinate is established.

A conditional MLP may be continuous in $x_i$ yet move the same value when unrelated fields change. Here interval index $k$, coefficient $a_i$, and $e_i^{\mathrm{loc}}$ depend only on the focal scalar and learned field parameters. Equality for identical values therefore holds by construction.

\subsection{Contextual Response Dynamics}

Concatenate the $N$ local tokens and $C$ categorical embeddings into $F=N+C$ field tokens and normalize each without an affine transform:
\begin{equation}
\bar e_f=\frac{e_f}{\sqrt{d^{-1}\lVert e_f\rVert_2^2+\epsilon_n}}.
\label{eq:rms_token}
\end{equation}
A separate head for each field maps $\bar e_f$ to an $m$-dimensional drive $\delta_f$ and gate $g_f$, thereby preserving field identity. Flattening gives $M=Fm$, vectors $\delta,\mathbf g$, and $\mathbf{s}^{(0)}=\tanh(\delta)$. For $t=0,\ldots,T-1$,
\begin{align}
\mathbf{u}^{(t)}&=(\mathbf{s}^{(t)}U)V,
&&U\in\mathbb{R}^{M\times r},\ V\in\mathbb{R}^{r\times M},
\label{eq:lowrank_message}\\
\widetilde{\mathbf{s}}^{(t+1)}
&=\tanh\!\left(\delta+\mathbf{g}\odot\mathbf{u}^{(t)}\right),
\nonumber\\
\mathbf{s}^{(t+1)}
&=(1-\alpha_t)\mathbf{s}^{(t)}+\alpha_t\widetilde{\mathbf{s}}^{(t+1)},
&&\alpha_t=\sigma(a_t).
\label{eq:state_update}
\end{align}
The shared $UV$ operator reduces communication from $O(M^2)$ to $O(Mr)$. Gates regulate transmission, bounded proposals limit amplitude, and fixed depth yields predictable cost. Context changes the response state, never the scalar coordinate.

The drive $\delta_f$ supplies local evidence, the gate $g_f$ controls incoming messages, and $UV$ supplies a shared reduced interaction subspace. Reintroducing $\delta$ prevents the state from becoming only a context summary. Separate heads retain field identity although the expensive operator is shared.

\paragraph{Bounded response.}
Since $\mathbf{s}^{(0)}$ and every proposal lie in $[-1,1]^M$ and $\alpha_t\in(0,1)$ makes each update convex, $\mathbf{s}^{(t)}\in[-1,1]^M$ for all $t$. This bound is independent of input scale, requires no iterative convergence test, and cannot alter the focal interval or interpolation coefficient.

\subsection{Numerical Readout that Preserves the Coordinate}

For numerical field $i$, the readout concatenates the retained stable coordinate, its gate, and all transient numerical states,
\begin{equation}
z_i=[\bar e_i;g_i;s_i^{(1)};\ldots;s_i^{(T)}],
\qquad e_i^{\num}=\operatorname{Readout}_i(z_i),
\label{eq:readout_features}
\end{equation}
where $\operatorname{Readout}_i$ is a SiLU head with two layers followed by a positive learned gain. Only the $N$ numerical outputs are returned. The frozen default is $d=16$, $K=16$, $m=8$, $r=16$, $T=3$, and head width 32.

Keeping $\bar e_i$ explicit prevents the response state from having to reconstruct location. The downstream task chooses how much of either path to use; Table~\ref{tab:ablation-suite} tests this division.

Only numerical readouts are returned. Thus improvements cannot arise from replacing categorical embeddings or inserting another interaction layer after all fields. Categorical tokens enter the consumer unchanged, while each numerical token carries its coordinate and contextual response.

\subsection{Complexity and Implementation Boundary}

For batch size $B$, interpolation costs $O(BNK)$ and response dynamics cost $O(BTMr)$; the remaining heads scale linearly with the number of fields. The module uses batched projections, two matrix multiplications per step, and pointwise nonlinearities. It requires no sample loop, matrix inverse, convergence test, or table that grows with vocabulary size. Section~\ref{sec:efficiency} measures complete training and inference paths; Appendix~\ref{app:method-details} gives implementation details.

\section{Experiments}

We ask whether \scalarlens transfers across datasets, backbones, and input scales (RQ1); whether the proposed separation of coordinate and response, together with its context sources, affects accuracy (RQ2); whether matched controls, known response functions, and sensitivity tests support that explanation (RQ3); and what balance between accuracy and computational cost can be achieved through implementation optimization (RQ4).

\subsection{Datasets, Backbones, and Shared Protocols}
\label{sec:protocol}

\begin{table}[t]
\caption{Dataset statistics. AutoML-A/E belong to one AutoML3 family; Criteo is a separate family.}
\label{tab:data}
\centering
\singlecoltableformat
\begin{tabular*}{\columnwidth}{@{\extracolsep{\fill}}lrrrrr@{}}
\toprule
Dataset & Num. & Cat. & Train & Valid & Test \\
\midrule
AutoML-A & 23 & 51 & 2.80M & 1.01M & 0.88M \\
AutoML-E & 6 & 25 & 5.14M & 1.82M & 1.78M \\
Criteo & 13 & 26 & 36.67M & 4.58M & 4.58M \\
\bottomrule
\end{tabular*}
\end{table}

AutoML-A/E come from AutoML3~\cite{guyon2019automl}; Criteo preserves all 13 numerical fields without a shared logarithmic transformation. We test nine backbones (DNN, DeepFM, WideDeep, NFM, PNN, DCN, DCNv2, AutoInt, and FiBiNET) with batch size 4096, 16-dimensional embeddings, Adam at $10^{-3}$, hidden widths $[256,128,64]$, no batch normalization, dropout, or regularization, global shuffling, early stopping based on validation AUC with patience 2, and seeds 2026--2028.

Every method returns one vector per numerical field and uses the same categorical path and backbone. Our primary protocol retains observed units and ranges, matching a serving path without separately materialized normalized features. Each cell is the mean over three seeds. Model selection uses validation AUC, whereas Tables~\ref{tab:full-results}--\ref{tab:ablation-suite} report test AUC for the selected checkpoint. Ranks use unrounded means, and Logloss provides a complementary measure of calibration.

\begin{table*}[t]
\caption{Complete frozen test results for 19 numerical representations. Means are truncated to four decimals; AUC subscripts in muted blue show standard deviations over three seeds. Every backbone has separate AUC and Logloss (LL) columns; bold and underline mark the two highest untruncated AUCs. Global Rank is the mean test AUC rank over 27 settings, and Wins counts the settings ranked first. To avoid repetition, both are shown only in the final dataset block. Published method sources are cited at first occurrence. ContextFiLM is an ablation and is excluded.}
\label{tab:full-results}
\centering
\tiny
\setlength{\tabcolsep}{0.8pt}
\renewcommand{\arraystretch}{0.80}
\resizebox{\textwidth}{!}{%
\begin{tabular}{lcccccccccccccccccccccc}
\toprule
\multirow{2}{*}{Method} & \multicolumn{2}{c}{DNN} & \multicolumn{2}{c}{DeepFM} & \multicolumn{2}{c}{WideDeep} & \multicolumn{2}{c}{NFM} & \multicolumn{2}{c}{PNN} & \multicolumn{2}{c}{DCN} & \multicolumn{2}{c}{DCNv2} & \multicolumn{2}{c}{AutoInt} & \multicolumn{2}{c}{FiBiNET} & \multicolumn{2}{c}{Mean} & \multicolumn{2}{c}{Global} \\
\cmidrule(lr){2-3} \cmidrule(lr){4-5} \cmidrule(lr){6-7} \cmidrule(lr){8-9} \cmidrule(lr){10-11} \cmidrule(lr){12-13} \cmidrule(lr){14-15} \cmidrule(lr){16-17} \cmidrule(lr){18-19} \cmidrule(lr){20-21} \cmidrule(lr){22-23}
 & AUC & LL & AUC & LL & AUC & LL & AUC & LL & AUC & LL & AUC & LL & AUC & LL & AUC & LL & AUC & LL & AUC & LL & Rank & Wins \\
\midrule
\multicolumn{23}{l}{\textbf{AutoML-A}} \\
Field & \aucstd{0.6671}{.0011} & 0.3019 & \aucstd{0.5568}{.0498} & 3.3910 & \aucstd{0.6679}{.0047} & 0.3789 & \aucstd{0.6666}{.0157} & 0.2487 & \aucstd{0.6541}{.0050} & 0.3089 & \aucstd{0.6592}{.0079} & 0.7683 & \aucstd{0.6008}{.0531} & 1.0023 & \aucstd{0.6585}{.0065} & 0.4435 & \aucstd{0.6052}{.0467} & 4.1776 & \aucstd{0.6373}{.0179} & 1.2246 & -- & -- \\
Linear & \aucstd{0.6073}{.0138} & 0.3906 & \aucstd{0.4944}{.0031} & 2.5562 & \aucstd{0.5572}{.0517} & 1.1676 & \aucstd{0.5087}{.0156} & 6.9871 & \aucstd{0.5828}{.0181} & 0.2778 & \aucstd{0.5040}{.0174} & 13.2011 & \aucstd{0.4972}{.0044} & 2.3688 & \aucstd{0.6555}{.0075} & 0.2454 & \aucstd{0.5798}{.0712} & 0.9240 & \aucstd{0.5541}{.0098} & 3.1243 & -- & -- \\
Linear-ReLU & \aucstd{0.6411}{.0169} & 0.2566 & \aucstd{0.5000}{.0000} & 34.0402 & \aucstd{0.5802}{.0323} & 2.9301 & \aucstd{0.5336}{.0589} & 1.7749 & \aucstd{0.6709}{.0009} & 0.2732 & \aucstd{0.5477}{.0285} & 4.9770 & \aucstd{0.6802}{.0129} & 0.7633 & \aucstd{0.6444}{.0174} & 0.2639 & \aucstd{0.5798}{.0564} & 2.3837 & \aucstd{0.5975}{.0062} & 5.2959 & -- & -- \\
Quantile bucket & \aucstd{0.7008}{.0024} & 0.2536 & \aucstd{0.5000}{.0000} & 2.0033 & \aucstd{0.6930}{.0084} & 0.2478 & \aucstd{0.6856}{.0109} & 0.2093 & \aucstd{0.7033}{.0036} & 0.2422 & \aucstd{0.7038}{.0050} & 0.2401 & \aucstd{0.7038}{.0030} & 0.2398 & \aucstd{0.7001}{.0021} & 0.2327 & \aucstd{0.6912}{.0049} & 0.2269 & \aucstd{0.6757}{.0031} & 0.4329 & -- & -- \\
Log-squared bucket & \aucstd{0.6825}{.0009} & 0.2463 & \aucstd{0.6848}{.0016} & 0.2413 & \aucstd{0.6820}{.0012} & 0.2439 & \aucstd{0.6680}{.0037} & 0.2300 & \aucstd{0.6832}{.0011} & 0.2493 & \aucstd{0.6800}{.0044} & 0.2486 & \aucstd{0.6805}{.0011} & 0.2508 & \aucstd{0.6869}{.0022} & 0.2259 & \aucstd{0.6796}{.0024} & 0.2505 & \aucstd{0.6808}{.0008} & 0.2430 & -- & -- \\
AutoDis~\cite{guo2021autodis} & \aucstd{0.6866}{.0014} & 0.2475 & \aucstd{0.6857}{.0074} & 0.2446 & \aucstd{0.6831}{.0043} & 0.2471 & \aucstd{0.6725}{.0017} & 0.2355 & \aucstd{0.6875}{.0022} & 0.2459 & \aucstd{0.6875}{.0046} & 0.2520 & \aucstd{0.6828}{.0022} & 0.2524 & \aucstd{0.7067}{.0039} & 0.2136 & \aucstd{0.6902}{.0055} & 0.2258 & \aucstd{0.6870}{.0015} & 0.2405 & -- & -- \\
PLE-A~\cite{gorishniy2022numemb} & \aucstd{\underline{0.7084}}{.0003} & 0.2428 & \aucstd{0.6740}{.0086} & 0.2353 & \aucstd{0.7011}{.0043} & 0.2474 & \aucstd{0.6890}{.0048} & 0.2162 & \aucstd{\underline{0.7102}}{.0004} & 0.2350 & \aucstd{0.7058}{.0016} & 0.2403 & \aucstd{0.7048}{.0013} & 0.2307 & \aucstd{0.7077}{.0041} & 0.2403 & \aucstd{0.7043}{.0037} & 0.2332 & \aucstd{0.7006}{.0020} & 0.2357 & -- & -- \\
PLE-B~\cite{gorishniy2025tabm} & \aucstd{0.6448}{.0233} & 0.2812 & \aucstd{0.4950}{.0028} & 2.6882 & \aucstd{0.5494}{.0561} & 7.2782 & \aucstd{0.5422}{.0438} & 4.0852 & \aucstd{0.6106}{.0467} & 0.2698 & \aucstd{0.4945}{.0068} & 2.1771 & \aucstd{0.4936}{.0051} & 2.7890 & \aucstd{0.6249}{.0056} & 0.2975 & \aucstd{0.5910}{.0790} & 0.9182 & \aucstd{0.5607}{.0118} & 2.3094 & -- & -- \\
PLE-ReLU~\cite{gorishniy2022numemb} & \aucstd{0.6217}{.0140} & 0.3168 & \aucstd{0.4941}{.0026} & 2.6659 & \aucstd{0.5642}{.0704} & 1.5567 & \aucstd{0.5823}{.0802} & 0.8843 & \aucstd{0.5905}{.0223} & 0.3224 & \aucstd{0.4951}{.0051} & 2.1926 & \aucstd{0.4929}{.0059} & 2.8894 & \aucstd{0.6287}{.0145} & 0.2466 & \aucstd{0.5787}{.0704} & 1.0117 & \aucstd{0.5609}{.0161} & 1.3429 & -- & -- \\
Periodic~\cite{gorishniy2022numemb} & \aucstd{0.6851}{.0019} & 0.2505 & \aucstd{0.6732}{.0026} & 0.2459 & \aucstd{0.6881}{.0018} & 0.2432 & \aucstd{0.6887}{.0014} & 0.2102 & \aucstd{0.6850}{.0002} & 0.2434 & \aucstd{0.6870}{.0018} & 0.2458 & \aucstd{0.6952}{.0026} & 0.2365 & \aucstd{0.6944}{.0016} & 0.2489 & \aucstd{0.6918}{.0062} & 0.2290 & \aucstd{0.6876}{.0009} & 0.2393 & -- & -- \\
PLR~\cite{gorishniy2022numemb} & \aucstd{0.6814}{.0013} & 0.2576 & \aucstd{0.5000}{.0000} & 34.0402 & \aucstd{0.6910}{.0021} & 0.2386 & \aucstd{0.6853}{.0025} & 0.2072 & \aucstd{0.6836}{.0008} & 0.2467 & \aucstd{0.6845}{.0021} & 0.2456 & \aucstd{0.6872}{.0072} & 0.2399 & \aucstd{0.6914}{.0057} & 0.2434 & \aucstd{0.6992}{.0036} & 0.2221 & \aucstd{0.6670}{.0014} & 3.9935 & -- & -- \\
B-spline~\cite{shtoff2024basis} & \aucstd{0.7072}{.0022} & 0.2463 & \aucstd{\underline{0.7105}}{.0011} & 0.2336 & \aucstd{\underline{0.7095}}{.0030} & 0.2351 & \aucstd{0.6897}{.0020} & 0.2230 & \aucstd{0.7068}{.0015} & 0.2371 & \aucstd{\underline{0.7074}}{.0018} & 0.2391 & \aucstd{\underline{0.7083}}{.0020} & 0.2391 & \aucstd{0.7115}{.0002} & 0.2207 & \aucstd{0.7033}{.0016} & 0.2370 & \aucstd{0.7060}{.0006} & 0.2346 & -- & -- \\
NaryDis-Lite~\cite{chen2022narydis} & \aucstd{0.7024}{.0019} & 0.2389 & \aucstd{0.6991}{.0030} & 0.2412 & \aucstd{0.6999}{.0026} & 0.2413 & \aucstd{0.6799}{.0019} & 0.2258 & \aucstd{0.7009}{.0032} & 0.2522 & \aucstd{0.6995}{.0030} & 0.2486 & \aucstd{0.7005}{.0011} & 0.2428 & \aucstd{0.7119}{.0030} & 0.2161 & \aucstd{0.6988}{.0010} & 0.2420 & \aucstd{0.6992}{.0005} & 0.2388 & -- & -- \\
NaryDis~\cite{chen2022narydis} & \aucstd{0.7043}{.0008} & 0.2418 & \aucstd{0.7043}{.0029} & 0.2351 & \aucstd{0.7034}{.0013} & 0.2368 & \aucstd{0.6890}{.0017} & 0.2253 & \aucstd{0.7034}{.0004} & 0.2433 & \aucstd{0.7036}{.0041} & 0.2435 & \aucstd{0.7045}{.0014} & 0.2447 & \aucstd{0.7108}{.0039} & 0.2166 & \aucstd{0.7024}{.0027} & 0.2530 & \aucstd{0.7028}{.0012} & 0.2378 & -- & -- \\
DEER~\cite{cheng2022deer} & \aucstd{0.7084}{.0012} & 0.2421 & \aucstd{0.7096}{.0007} & 0.2408 & \aucstd{0.7078}{.0003} & 0.2428 & \aucstd{\underline{0.6918}}{.0033} & 0.2259 & \aucstd{0.7080}{.0020} & 0.2417 & \aucstd{0.7070}{.0008} & 0.2395 & \aucstd{0.7073}{.0022} & 0.2391 & \aucstd{\textbf{0.7158}}{.0018} & 0.2219 & \aucstd{0.7062}{.0013} & 0.2486 & \aucstd{\underline{0.7069}}{.0005} & 0.2380 & -- & -- \\
DAE~\cite{shen2023dae} & \aucstd{0.7012}{.0013} & 0.2363 & \aucstd{0.7019}{.0017} & 0.2377 & \aucstd{0.6995}{.0034} & 0.2392 & \aucstd{0.6871}{.0021} & 0.2281 & \aucstd{0.6996}{.0013} & 0.2443 & \aucstd{0.6983}{.0029} & 0.2455 & \aucstd{0.7007}{.0034} & 0.2407 & \aucstd{0.7059}{.0024} & 0.2214 & \aucstd{0.6940}{.0064} & 0.2426 & \aucstd{0.6987}{.0012} & 0.2373 & -- & -- \\
DAES-Gate~\cite{liu2026daes} & \aucstd{0.7057}{.0014} & 0.2437 & \aucstd{0.7035}{.0046} & 0.2397 & \aucstd{0.7061}{.0027} & 0.2386 & \aucstd{0.6886}{.0083} & 0.2223 & \aucstd{0.7038}{.0019} & 0.2487 & \aucstd{0.7035}{.0027} & 0.2417 & \aucstd{0.7064}{.0035} & 0.2389 & \aucstd{0.7111}{.0016} & 0.2258 & \aucstd{\underline{0.7074}}{.0049} & 0.2367 & \aucstd{0.7040}{.0003} & 0.2373 & -- & -- \\
DAES-Tran~\cite{liu2026daes} & \aucstd{0.7024}{.0033} & 0.2503 & \aucstd{0.7008}{.0052} & 0.2355 & \aucstd{0.7007}{.0049} & 0.2422 & \aucstd{0.6845}{.0066} & 0.2289 & \aucstd{0.7050}{.0038} & 0.2443 & \aucstd{0.7038}{.0016} & 0.2437 & \aucstd{0.7010}{.0035} & 0.2411 & \aucstd{0.7072}{.0041} & 0.2203 & \aucstd{0.7042}{.0038} & 0.2420 & \aucstd{0.7011}{.0005} & 0.2387 & -- & -- \\
\textbf{ScalarLens} & \aucstd{\textbf{0.7123}}{.0021} & 0.2270 & \aucstd{\textbf{0.7158}}{.0015} & 0.2200 & \aucstd{\textbf{0.7151}}{.0021} & 0.2239 & \aucstd{\textbf{0.7125}}{.0016} & 0.2217 & \aucstd{\textbf{0.7150}}{.0025} & 0.2273 & \aucstd{\textbf{0.7135}}{.0008} & 0.2277 & \aucstd{\textbf{0.7088}}{.0033} & 0.2276 & \aucstd{\underline{0.7153}}{.0024} & 0.2174 & \aucstd{\textbf{0.7172}}{.0015} & 0.2189 & \aucstd{\textbf{0.7139}}{.0006} & 0.2235 & -- & -- \\
\midrule
\multicolumn{23}{l}{\textbf{AutoML-E}} \\
Field & \aucstd{0.6103}{.1911} & 0.4041 & \aucstd{0.5000}{.0000} & 0.5703 & \aucstd{0.5000}{.0000} & 12.2046 & \aucstd{0.5000}{.0000} & 0.5703 & \aucstd{0.5000}{.0000} & 0.5703 & \aucstd{0.4999}{.0002} & 12.2187 & \aucstd{0.5001}{.0002} & 23.8255 & \aucstd{0.7114}{.1831} & 0.2367 & \aucstd{0.5000}{.0000} & 35.4732 & \aucstd{0.5357}{.0359} & 9.5638 & -- & -- \\
Linear & \aucstd{0.5000}{.0000} & 23.8389 & \aucstd{0.5000}{.0000} & 0.5703 & \aucstd{0.4995}{.0009} & 23.8171 & \aucstd{0.4938}{.0106} & 13.5131 & \aucstd{0.5038}{.0091} & 12.9760 & \aucstd{0.5022}{.0147} & 20.2365 & \aucstd{0.4975}{.0030} & 12.6317 & \aucstd{0.5000}{.0000} & 0.5703 & \aucstd{0.4999}{.0000} & 0.5714 & \aucstd{0.4996}{.0034} & 12.0806 & -- & -- \\
Linear-ReLU & \aucstd{0.5000}{.0000} & 23.8389 & \aucstd{0.5000}{.0000} & 35.4732 & \aucstd{0.5000}{.0002} & 0.5770 & \aucstd{0.4941}{.0102} & 14.6022 & \aucstd{0.4948}{.0076} & 22.6233 & \aucstd{0.5018}{.0036} & 1.6626 & \aucstd{0.5052}{.0131} & 17.1688 & \aucstd{0.5000}{.0000} & 0.5703 & \aucstd{0.5007}{.0013} & 23.6975 & \aucstd{0.4996}{.0025} & 15.5793 & -- & -- \\
Quantile bucket & \aucstd{0.8360}{.0008} & 0.0706 & \aucstd{0.7838}{.0012} & 0.1149 & \aucstd{0.8345}{.0005} & 0.0712 & \aucstd{0.8266}{.0027} & 0.0698 & \aucstd{0.8366}{.0007} & 0.0715 & \aucstd{0.8354}{.0007} & 0.0719 & \aucstd{0.8358}{.0011} & 0.0717 & \aucstd{0.8345}{.0004} & 0.0713 & \aucstd{0.8343}{.0012} & 0.0712 & \aucstd{0.8286}{.0009} & 0.0760 & -- & -- \\
Log-squared bucket & \aucstd{0.8379}{.0016} & 0.0698 & \aucstd{0.8383}{.0031} & 0.0708 & \aucstd{0.8375}{.0033} & 0.0710 & \aucstd{0.8387}{.0003} & 0.0703 & \aucstd{0.8383}{.0013} & 0.0705 & \aucstd{0.8359}{.0031} & 0.0702 & \aucstd{0.8365}{.0008} & 0.0713 & \aucstd{0.8415}{.0009} & 0.0690 & \aucstd{0.8340}{.0012} & 0.0721 & \aucstd{0.8376}{.0007} & 0.0706 & -- & -- \\
AutoDis & \aucstd{0.8361}{.0010} & 0.0713 & \aucstd{0.8379}{.0011} & 0.0700 & \aucstd{0.8367}{.0011} & 0.0702 & \aucstd{0.8381}{.0010} & 0.0715 & \aucstd{0.8361}{.0018} & 0.0705 & \aucstd{0.8357}{.0008} & 0.0709 & \aucstd{0.8356}{.0020} & 0.0704 & \aucstd{0.8417}{.0009} & 0.0692 & \aucstd{0.8320}{.0015} & 0.0728 & \aucstd{0.8367}{.0004} & 0.0708 & -- & -- \\
PLE-A & \aucstd{0.8387}{.0009} & 0.0713 & \aucstd{0.8388}{.0006} & 0.0707 & \aucstd{0.8378}{.0019} & 0.0716 & \aucstd{0.8373}{.0012} & 0.0705 & \aucstd{0.8394}{.0020} & 0.0715 & \aucstd{0.8389}{.0006} & 0.0710 & \aucstd{0.8393}{.0009} & 0.0718 & \aucstd{0.8399}{.0005} & 0.0702 & \aucstd{0.8363}{.0008} & 0.0721 & \aucstd{0.8385}{.0004} & 0.0712 & -- & -- \\
PLE-B & \aucstd{0.5000}{.0000} & 12.2046 & \aucstd{0.5000}{.0000} & 0.5703 & \aucstd{0.5001}{.0003} & 0.5971 & \aucstd{0.4955}{.0077} & 23.3506 & \aucstd{0.5041}{.0075} & 12.7098 & \aucstd{0.4949}{.0053} & 17.7530 & \aucstd{0.4961}{.0211} & 19.5144 & \aucstd{0.5000}{.0000} & 12.2046 & \aucstd{0.4950}{.0047} & 11.9498 & \aucstd{0.4984}{.0032} & 12.3171 & -- & -- \\
PLE-ReLU & \aucstd{0.5000}{.0000} & 12.2046 & \aucstd{0.5000}{.0000} & 0.5703 & \aucstd{0.5001}{.0003} & 0.5971 & \aucstd{0.4955}{.0077} & 23.3506 & \aucstd{0.5041}{.0075} & 12.7098 & \aucstd{0.4949}{.0053} & 17.7530 & \aucstd{0.4961}{.0211} & 19.5144 & \aucstd{0.5000}{.0000} & 12.2046 & \aucstd{0.4950}{.0047} & 11.9498 & \aucstd{0.4984}{.0032} & 12.3171 & -- & -- \\
Periodic & \aucstd{0.8376}{.0017} & 0.0718 & \aucstd{0.8379}{.0008} & 0.0695 & \aucstd{0.8392}{.0005} & 0.0701 & \aucstd{0.8373}{.0006} & 0.0696 & \aucstd{0.8388}{.0004} & 0.0711 & \aucstd{0.8388}{.0002} & 0.0696 & \aucstd{0.8384}{.0013} & 0.0703 & \aucstd{0.8370}{.0002} & 0.0710 & \aucstd{0.8386}{.0016} & 0.0710 & \aucstd{0.8382}{.0002} & 0.0704 & -- & -- \\
PLR & \aucstd{0.8386}{.0015} & 0.0713 & \aucstd{0.8372}{.0006} & 0.0697 & \aucstd{\underline{0.8404}}{.0013} & 0.0696 & \aucstd{0.8381}{.0003} & 0.0694 & \aucstd{0.8394}{.0004} & 0.0716 & \aucstd{\underline{0.8392}}{.0005} & 0.0695 & \aucstd{\underline{0.8395}}{.0019} & 0.0705 & \aucstd{0.8370}{.0008} & 0.0714 & \aucstd{\underline{0.8396}}{.0003} & 0.0706 & \aucstd{\underline{0.8388}}{.0003} & 0.0704 & -- & -- \\
B-spline & \aucstd{\underline{0.8387}}{.0019} & 0.0703 & \aucstd{\underline{0.8398}}{.0005} & 0.0701 & \aucstd{0.8393}{.0008} & 0.0701 & \aucstd{0.8389}{.0004} & 0.0697 & \aucstd{0.8364}{.0020} & 0.0713 & \aucstd{0.8378}{.0023} & 0.0700 & \aucstd{0.8357}{.0015} & 0.0699 & \aucstd{\textbf{0.8435}}{.0007} & 0.0697 & \aucstd{0.8298}{.0054} & 0.0743 & \aucstd{0.8377}{.0009} & 0.0706 & -- & -- \\
NaryDis-Lite & \aucstd{0.8374}{.0015} & 0.0711 & \aucstd{0.8371}{.0015} & 0.0712 & \aucstd{0.8366}{.0008} & 0.0709 & \aucstd{0.8385}{.0005} & 0.0713 & \aucstd{0.8374}{.0012} & 0.0714 & \aucstd{0.8356}{.0021} & 0.0725 & \aucstd{0.8338}{.0012} & 0.0722 & \aucstd{0.8415}{.0012} & 0.0685 & \aucstd{0.8328}{.0008} & 0.0720 & \aucstd{0.8367}{.0001} & 0.0712 & -- & -- \\
NaryDis & \aucstd{0.8354}{.0019} & 0.0726 & \aucstd{0.8374}{.0012} & 0.0703 & \aucstd{0.8367}{.0014} & 0.0705 & \aucstd{0.8386}{.0005} & 0.0709 & \aucstd{0.8373}{.0019} & 0.0704 & \aucstd{0.8349}{.0033} & 0.0707 & \aucstd{0.8363}{.0012} & 0.0707 & \aucstd{0.8427}{.0005} & 0.0691 & \aucstd{0.8321}{.0026} & 0.0738 & \aucstd{0.8368}{.0003} & 0.0710 & -- & -- \\
DEER & \aucstd{0.8373}{.0010} & 0.0710 & \aucstd{0.8388}{.0017} & 0.0697 & \aucstd{0.8381}{.0015} & 0.0699 & \aucstd{\underline{0.8398}}{.0004} & 0.0707 & \aucstd{0.8377}{.0021} & 0.0707 & \aucstd{0.8373}{.0033} & 0.0716 & \aucstd{0.8374}{.0016} & 0.0707 & \aucstd{0.8429}{.0015} & 0.0682 & \aucstd{0.8360}{.0031} & 0.0720 & \aucstd{0.8384}{.0006} & 0.0705 & -- & -- \\
DAE & \aucstd{0.8373}{.0011} & 0.0699 & \aucstd{0.8380}{.0014} & 0.0699 & \aucstd{0.8372}{.0011} & 0.0699 & \aucstd{0.8387}{.0003} & 0.0706 & \aucstd{\underline{0.8395}}{.0009} & 0.0699 & \aucstd{0.8365}{.0017} & 0.0710 & \aucstd{0.8372}{.0005} & 0.0704 & \aucstd{0.8432}{.0004} & 0.0684 & \aucstd{0.8324}{.0029} & 0.0740 & \aucstd{0.8378}{.0004} & 0.0704 & -- & -- \\
DAES-Gate & \aucstd{0.8386}{.0014} & 0.0706 & \aucstd{0.8352}{.0026} & 0.0720 & \aucstd{0.8344}{.0026} & 0.0721 & \aucstd{0.8389}{.0014} & 0.0711 & \aucstd{0.8393}{.0013} & 0.0708 & \aucstd{0.8375}{.0024} & 0.0703 & \aucstd{0.8369}{.0013} & 0.0710 & \aucstd{0.8423}{.0018} & 0.0688 & \aucstd{0.8308}{.0018} & 0.0738 & \aucstd{0.8371}{.0002} & 0.0712 & -- & -- \\
DAES-Tran & \aucstd{0.8376}{.0015} & 0.0716 & \aucstd{0.8360}{.0058} & 0.0717 & \aucstd{0.8356}{.0060} & 0.0713 & \aucstd{0.8384}{.0007} & 0.0703 & \aucstd{0.8344}{.0022} & 0.0734 & \aucstd{0.8360}{.0023} & 0.0707 & \aucstd{0.8363}{.0019} & 0.0713 & \aucstd{0.8422}{.0003} & 0.0695 & \aucstd{0.8295}{.0055} & 0.0722 & \aucstd{0.8362}{.0011} & 0.0713 & -- & -- \\
\textbf{ScalarLens} & \aucstd{\textbf{0.8403}}{.0006} & 0.0699 & \aucstd{\textbf{0.8430}}{.0002} & 0.0687 & \aucstd{\textbf{0.8426}}{.0007} & 0.0690 & \aucstd{\textbf{0.8415}}{.0005} & 0.0689 & \aucstd{\textbf{0.8433}}{.0014} & 0.0692 & \aucstd{\textbf{0.8418}}{.0017} & 0.0698 & \aucstd{\textbf{0.8425}}{.0006} & 0.0699 & \aucstd{\underline{0.8435}}{.0005} & 0.0684 & \aucstd{\textbf{0.8441}}{.0005} & 0.0689 & \aucstd{\textbf{0.8425}}{.0004} & 0.0692 & -- & -- \\
\midrule
\multicolumn{23}{l}{\textbf{Criteo}} \\
Field & \aucstd{0.8059}{.0003} & 0.4459 & \aucstd{0.7931}{.0008} & 0.4580 & \aucstd{0.8025}{.0012} & 0.4495 & \aucstd{0.7978}{.0006} & 0.4546 & \aucstd{0.8079}{.0008} & 0.4438 & \aucstd{0.8029}{.0003} & 0.4485 & \aucstd{0.8044}{.0000} & 0.4485 & \aucstd{0.8055}{.0005} & 0.4461 & \aucstd{0.8054}{.0015} & 0.4488 & \aucstd{0.8028}{.0003} & 0.4493 & 15.259 & 0/27 \\
Linear & \aucstd{0.8053}{.0001} & 0.4463 & \aucstd{0.7886}{.0073} & 0.4692 & \aucstd{0.7833}{.0195} & 0.5241 & \aucstd{0.7848}{.0032} & 0.4703 & \aucstd{0.8011}{.0011} & 0.4511 & \aucstd{0.7979}{.0041} & 0.4526 & \aucstd{0.8029}{.0013} & 0.4502 & \aucstd{0.8026}{.0005} & 0.4491 & \aucstd{0.7822}{.0198} & 0.5714 & \aucstd{0.7943}{.0026} & 0.4761 & 17.556 & 0/27 \\
Linear-ReLU & \aucstd{0.8079}{.0002} & 0.4439 & \aucstd{0.8023}{.0002} & 0.4499 & \aucstd{0.7964}{.0014} & 0.4540 & \aucstd{0.7738}{.0248} & 0.5875 & \aucstd{0.8097}{.0003} & 0.4426 & \aucstd{0.8075}{.0003} & 0.4440 & \aucstd{0.8108}{.0002} & 0.4412 & \aucstd{0.8036}{.0008} & 0.4478 & \aucstd{0.7937}{.0056} & 0.4681 & \aucstd{0.8006}{.0026} & 0.4643 & 15.444 & 0/27 \\
Quantile bucket & \aucstd{0.8083}{.0001} & 0.4435 & \aucstd{0.7927}{.0021} & 0.4601 & \aucstd{0.8073}{.0002} & 0.4445 & \aucstd{0.7992}{.0001} & 0.4523 & \aucstd{0.8085}{.0002} & 0.4430 & \aucstd{0.8082}{.0001} & 0.4437 & \aucstd{0.8107}{.0001} & 0.4410 & \aucstd{0.8082}{.0001} & 0.4434 & \aucstd{0.8095}{.0002} & 0.4421 & \aucstd{0.8059}{.0003} & 0.4460 & 10.815 & 0/27 \\
Log-squared bucket & \aucstd{0.8096}{.0001} & 0.4423 & \aucstd{0.8068}{.0002} & 0.4449 & \aucstd{0.8090}{.0001} & 0.4432 & \aucstd{0.8042}{.0002} & 0.4473 & \aucstd{0.8108}{.0001} & 0.4409 & \aucstd{0.8097}{.0001} & 0.4423 & \aucstd{0.8112}{.0001} & 0.4408 & \aucstd{0.8098}{.0002} & 0.4420 & \aucstd{0.8103}{.0001} & 0.4419 & \aucstd{0.8090}{.0001} & 0.4428 & 9.519 & 0/27 \\
AutoDis & \aucstd{0.8078}{.0010} & 0.4436 & \aucstd{0.8048}{.0011} & 0.4464 & \aucstd{0.8067}{.0009} & 0.4448 & \aucstd{0.8020}{.0009} & 0.4499 & \aucstd{0.8078}{.0013} & 0.4444 & \aucstd{0.8069}{.0012} & 0.4446 & \aucstd{0.8087}{.0017} & 0.4429 & \aucstd{0.8078}{.0007} & 0.4438 & \aucstd{0.8075}{.0012} & 0.4444 & \aucstd{0.8067}{.0011} & 0.4450 & 11.148 & 0/27 \\
PLE-A & \aucstd{0.8104}{.0002} & 0.4416 & \aucstd{0.8066}{.0004} & 0.4450 & \aucstd{0.8097}{.0001} & 0.4422 & \aucstd{0.8050}{.0004} & 0.4468 & \aucstd{0.8113}{.0000} & 0.4407 & \aucstd{0.8102}{.0001} & 0.4420 & \aucstd{0.8121}{.0001} & 0.4399 & \aucstd{0.8104}{.0000} & 0.4413 & \aucstd{\underline{0.8116}}{.0002} & 0.4402 & \aucstd{0.8097}{.0000} & 0.4422 & 5.370 & 0/27 \\
PLE-B & \aucstd{0.8067}{.0002} & 0.4449 & \aucstd{0.7971}{.0001} & 0.4544 & \aucstd{0.7928}{.0064} & 0.4641 & \aucstd{0.7791}{.0227} & 0.5562 & \aucstd{0.8025}{.0004} & 0.4490 & \aucstd{0.8050}{.0002} & 0.4471 & \aucstd{0.8055}{.0008} & 0.4478 & \aucstd{0.8040}{.0004} & 0.4476 & \aucstd{0.7969}{.0017} & 0.4547 & \aucstd{0.7988}{.0020} & 0.4629 & 16.704 & 0/27 \\
PLE-ReLU & \aucstd{0.8050}{.0007} & 0.4466 & \aucstd{0.7926}{.0016} & 0.4596 & \aucstd{0.7761}{.0307} & 0.5910 & \aucstd{0.7832}{.0024} & 0.4772 & \aucstd{0.8014}{.0009} & 0.4507 & \aucstd{0.8009}{.0021} & 0.4509 & \aucstd{0.8034}{.0006} & 0.4496 & \aucstd{0.8034}{.0009} & 0.4481 & \aucstd{0.7710}{.0377} & 0.6955 & \aucstd{0.7930}{.0034} & 0.4966 & 17.926 & 0/27 \\
Periodic & \aucstd{0.8062}{.0004} & 0.4452 & \aucstd{0.8028}{.0015} & 0.4490 & \aucstd{0.8051}{.0008} & 0.4461 & \aucstd{0.8011}{.0004} & 0.4509 & \aucstd{0.8073}{.0002} & 0.4441 & \aucstd{0.8059}{.0007} & 0.4456 & \aucstd{0.8085}{.0003} & 0.4430 & \aucstd{0.8057}{.0003} & 0.4454 & \aucstd{0.8076}{.0003} & 0.4440 & \aucstd{0.8056}{.0003} & 0.4459 & 10.259 & 0/27 \\
PLR & \aucstd{0.8065}{.0002} & 0.4449 & \aucstd{0.8024}{.0007} & 0.4495 & \aucstd{0.8054}{.0003} & 0.4459 & \aucstd{0.8003}{.0004} & 0.4506 & \aucstd{0.8077}{.0005} & 0.4439 & \aucstd{0.8065}{.0005} & 0.4448 & \aucstd{0.8083}{.0002} & 0.4432 & \aucstd{0.8068}{.0002} & 0.4444 & \aucstd{0.8076}{.0004} & 0.4437 & \aucstd{0.8057}{.0001} & 0.4457 & 10.074 & 0/27 \\
B-spline & \aucstd{0.8078}{.0002} & 0.4440 & \aucstd{0.8052}{.0003} & 0.4461 & \aucstd{0.8074}{.0003} & 0.4445 & \aucstd{0.7985}{.0002} & 0.4528 & \aucstd{0.8082}{.0002} & 0.4436 & \aucstd{0.8080}{.0003} & 0.4439 & \aucstd{0.8091}{.0001} & 0.4430 & \aucstd{0.8083}{.0003} & 0.4435 & \aucstd{0.8091}{.0002} & 0.4430 & \aucstd{0.8068}{.0001} & 0.4449 & 6.481 & 1/27 \\
NaryDis-Lite & \aucstd{0.8104}{.0001} & 0.4413 & \aucstd{0.8079}{.0001} & 0.4437 & \aucstd{0.8098}{.0001} & 0.4423 & \aucstd{0.8045}{.0003} & 0.4472 & \aucstd{0.8113}{.0001} & 0.4404 & \aucstd{0.8103}{.0002} & 0.4416 & \aucstd{0.8120}{.0001} & 0.4401 & \aucstd{0.8107}{.0001} & 0.4410 & \aucstd{0.8109}{.0001} & 0.4413 & \aucstd{0.8098}{.0000} & 0.4421 & 7.926 & 0/27 \\
NaryDis & \aucstd{0.8107}{.0001} & 0.4411 & \aucstd{0.8080}{.0001} & 0.4437 & \aucstd{\underline{0.8100}}{.0001} & 0.4421 & \aucstd{0.8043}{.0002} & 0.4474 & \aucstd{0.8114}{.0001} & 0.4404 & \aucstd{\underline{0.8107}}{.0001} & 0.4414 & \aucstd{0.8121}{.0001} & 0.4398 & \aucstd{\underline{0.8110}}{.0001} & 0.4410 & \aucstd{0.8110}{.0000} & 0.4412 & \aucstd{0.8099}{.0001} & 0.4420 & 6.407 & 0/27 \\
DEER & \aucstd{\underline{0.8107}}{.0001} & 0.4412 & \aucstd{\underline{0.8082}}{.0001} & 0.4435 & \aucstd{0.8100}{.0001} & 0.4420 & \aucstd{0.8052}{.0001} & 0.4466 & \aucstd{\underline{0.8117}}{.0000} & 0.4401 & \aucstd{0.8106}{.0002} & 0.4416 & \aucstd{0.8122}{.0001} & 0.4398 & \aucstd{0.8109}{.0001} & 0.4409 & \aucstd{0.8110}{.0001} & 0.4409 & \aucstd{\underline{0.8101}}{.0000} & 0.4419 & 3.704 & 1/27 \\
DAE & \aucstd{0.7970}{.0013} & 0.4536 & \aucstd{0.7966}{.0010} & 0.4546 & \aucstd{0.7967}{.0022} & 0.4542 & \aucstd{0.7950}{.0022} & 0.4557 & \aucstd{0.8000}{.0022} & 0.4510 & \aucstd{0.7985}{.0027} & 0.4525 & \aucstd{0.8018}{.0021} & 0.4493 & \aucstd{0.7998}{.0020} & 0.4514 & \aucstd{0.7994}{.0020} & 0.4518 & \aucstd{0.7983}{.0018} & 0.4527 & 10.852 & 0/27 \\
DAES-Gate & \aucstd{0.8105}{.0002} & 0.4412 & \aucstd{0.8077}{.0001} & 0.4439 & \aucstd{0.8099}{.0002} & 0.4421 & \aucstd{\underline{0.8057}}{.0001} & 0.4463 & \aucstd{0.8116}{.0001} & 0.4401 & \aucstd{0.8105}{.0001} & 0.4415 & \aucstd{\underline{0.8122}}{.0001} & 0.4398 & \aucstd{0.8107}{.0001} & 0.4415 & \aucstd{0.8113}{.0001} & 0.4407 & \aucstd{0.8100}{.0001} & 0.4419 & 5.519 & 0/27 \\
DAES-Tran & \aucstd{0.8101}{.0001} & 0.4415 & \aucstd{0.8071}{.0001} & 0.4448 & \aucstd{0.8095}{.0003} & 0.4425 & \aucstd{0.8055}{.0001} & 0.4462 & \aucstd{0.8106}{.0001} & 0.4413 & \aucstd{0.8101}{.0000} & 0.4418 & \aucstd{0.8117}{.0002} & 0.4404 & \aucstd{0.8101}{.0001} & 0.4419 & \aucstd{0.8109}{.0000} & 0.4413 & \aucstd{0.8095}{.0000} & 0.4424 & 7.963 & 0/27 \\
\textbf{ScalarLens} & \aucstd{\textbf{0.8116}}{.0002} & 0.4404 & \aucstd{\textbf{0.8088}}{.0003} & 0.4431 & \aucstd{\textbf{0.8111}}{.0001} & 0.4408 & \aucstd{\textbf{0.8089}}{.0001} & 0.4430 & \aucstd{\textbf{0.8120}}{.0001} & 0.4397 & \aucstd{\textbf{0.8117}}{.0002} & 0.4404 & \aucstd{\textbf{0.8133}}{.0001} & 0.4387 & \aucstd{\textbf{0.8115}}{.0001} & 0.4403 & \aucstd{\textbf{0.8127}}{.0001} & 0.4394 & \aucstd{\textbf{0.8113}}{.0000} & 0.4407 & \textbf{1.074} & \textbf{25/27} \\
\bottomrule
\end{tabular}%
}
\end{table*}

\begin{table}[t]
\caption{\textbf{Primary comparison on original numerical scales and secondary audit under standardization.} Each pair is one combination of dataset and backbone averaged over the same three seeds ($n=27$). Intervals bootstrap paired settings; $p$ values are from Wilcoxon signed rank tests with Holm correction over all six comparisons. Positive values favor \scalarlens.}
\label{tab:preprocessing-robustness}
\centering
\singlecoltableformat
\begin{tabular*}{\columnwidth}{@{\extracolsep{\fill}}llcccc@{}}
\toprule
Input & Comparator & \makecell{Mean\\$\Delta$AUC} & \makecell{95\% bootstrap\\CI} & W/T/L & \makecell{Holm\\$p$} \\
\midrule
\multirow{3}{*}{Raw}
 & DEER          & +0.0041 & $[+0.0027,+0.0059]$ & 26/0/1 & $2.98\!\times\!10^{-7}$ \\
 & DAES          & +0.0055 & $[+0.0037,+0.0077]$ & 27/0/0 & $8.94\!\times\!10^{-8}$ \\
 & NaryDis  & +0.0060 & $[+0.0041,+0.0082]$ & 27/0/0 & $8.94\!\times\!10^{-8}$ \\
\midrule
\multirow{3}{*}{Z-score}
 & DEER          & +0.0031 & $[+0.0014,+0.0053]$ & 18/0/9 & $1.25\!\times\!10^{-3}$ \\
 & DAES          & +0.0047 & $[+0.0026,+0.0072]$ & 21/0/6 & $9.49\!\times\!10^{-5}$ \\
 & NaryDis  & +0.0103 & $[+0.0064,+0.0146]$ & 25/0/2 & $1.92\!\times\!10^{-6}$ \\
\bottomrule
\end{tabular*}
\end{table}

\begin{table*}[t]
\caption{Unified mechanism ablations on test AUC. The three blocks contain interventions on context sources, interventions on components, and matched alternatives. \scalarlens AUC is the mean over three seeds; every other entry is $10^3\!\times\Delta$AUC (variant minus \scalarlens), so negative values indicate degradation. All variants preserve the interface that returns only numerical embeddings.}
\label{tab:ablation-suite}
\centering
\scriptsize
\setlength{\tabcolsep}{2.0pt}
\renewcommand{\arraystretch}{1.08}
\resizebox{\textwidth}{!}{%
\begin{tabular}{*{13}{c}}
\toprule
& & & \multicolumn{3}{c}{Interventions on context sources} & \multicolumn{4}{c}{Component interventions} & \multicolumn{3}{c}{Matched alternatives} \\
\cmidrule(lr){4-6}\cmidrule(lr){7-10}\cmidrule(lr){11-13}
Dataset & Backbone & \shortstack{ScalarLens\\AUC} & \shortstack{Numerical context\\only} & \shortstack{Categorical context\\only} & \shortstack{Local\\only} & NoNorm & FrozenMessage & StopGrad & $T{=}1$ & DEER+Norm & LocalMLP & ContextFiLM \\
\midrule
AutoML-A & DCNv2  & 0.7088 & $-2.351$ & \textbf{$+1.509$} & $-2.194$ & $-3.083$ & \textbf{$+1.180$} & $-0.368$ & $-0.160$ & $-3.494$ & $-2.322$ & $-2.287$ \\
AutoML-A & DeepFM & 0.7158 & $-9.944$ & \textbf{$+0.164$} & $-9.089$ & $-8.305$ & $-1.682$ & $-7.595$ & $-4.326$ & $-9.770$ & $-10.086$ & $-0.592$ \\
AutoML-E & DCNv2  & 0.8425 & $-4.693$ & $-0.117$ & $-4.827$ & $-5.002$ & $-0.026$ & $-3.629$ & $-0.393$ & $-5.111$ & $-5.204$ & $-1.480$ \\
AutoML-E & DeepFM & 0.8430 & $-5.799$ & $-0.178$ & $-5.645$ & $-4.692$ & $-0.438$ & $-2.685$ & $-1.100$ & $-5.953$ & $-7.085$ & $-2.394$ \\
Criteo & DCNv2  & 0.8133 & $-0.428$ & $-0.119$ & $-0.646$ & $-0.829$ & $-0.004$ & $-0.543$ & $-0.352$ & $-0.945$ & $-0.772$ & \textbf{$+0.298$} \\
Criteo & DeepFM & 0.8088 & $-0.102$ & $-0.031$ & $-0.453$ & $-0.339$ & $-0.005$ & $-0.285$ & $-0.151$ & $-0.761$ & $-0.450$ & $-0.189$ \\
\bottomrule
\end{tabular}}
\end{table*}

As a secondary control, we rerun the complete matrix after applying one z-score transformation fitted on the training split and shared by all methods. This experiment asks whether the rankings survive conventional stabilization, rather than which method should own a production preprocessing pipeline. No method receives private preprocessing, a different split, or hyperparameters specific to a dataset. Each matrix contains $3\!\times\!9\!\times\!19\!\times\!3=1{,}539$ runs, so the conclusions do not depend on a selected subset of baselines.

\subsection{Comparison with Numerical Embedding Methods}
\label{sec:full-comparison}

Table~\ref{tab:full-results} compares 19 configurations. PLE-A, Periodic, and PLR use the official implementation~\cite{gorishniy2022numemb}; PLE-B follows TabM~\cite{gorishniy2025tabm}. DAE, DEER, NaryDis, and DAES-Gate/Tran are implementations based on the published descriptions because no official module compatible with our code version was available; the DAES rows exclude its streaming estimator. Linear branches receive the same 16-dimensional numerical token.

The comparison spans raw scalars, fixed discretization, learned local and spectral bases, distributional and hierarchical encoders, and conditional DAES controls. Implementations based on the papers preserve the reported mechanism rather than every original data pipeline; the shared codebase isolates the numerical representation.

At least one encoder invariant to context improves Field in every setting, but its identity changes: DEER leads nine, B-spline seven, PLR four, PLE-A and NaryDis three each, and DAE one. \scalarlens instead ranks first in 25 of 27 settings and second in the two AutoInt exceptions, with mean rank 1.074. It wins eight of nine backbones on each AutoML dataset and all nine on Criteo, exceeding the competitor with the strongest mean on each dataset by 0.0070, 0.0037, and 0.0012 AUC, respectively. Its mean Logloss is also lowest on all three datasets (0.2235/0.0692/0.4407), so the AUC gain does not systematically sacrifice probability quality.

DEER leads \scalarlens by 0.0005 on AutoML-A/AutoInt, and B-spline ties it at displayed precision on AutoML-E/AutoInt. All settings with DNN, factorization, product, and cross network backbones are wins. The pattern supports a transferable representation interface while stopping short of universal dominance.

\paragraph{Primary results on original numerical scales and secondary scale audit.}
Keeping the original scale makes range handling part of the representation task. This protocol is operationally relevant when numerical fields arrive from independently versioned sources: materialized standardization duplicates data and I/O, whereas online standardization adds a transformation and a synchronization dependency. The unstable AutoML cells are therefore valid failures of compatibility with the original numerical ranges, not tuned upper bounds.\footnote{Identical PLE-B and PLE-ReLU entries on AutoML-E come from independent jobs. Values as large as $1.46\!\times\!10^{7}$ saturate both methods.}

The audit under shared z-score standardization removes all 60 cells below 0.55 AUC. \scalarlens nevertheless remains ahead of DEER, DAES, and NaryDis by $+0.0031$, $+0.0047$, and $+0.0103$ mean AUC (Table~\ref{tab:preprocessing-robustness}). All paired comparisons remain significant after Holm correction. Compatibility with original numerical scales therefore explains part, but not all, of the gain. We use this audit to test that alternative explanation rather than as the primary deployment result.

Beyond Figure~\ref{fig:motivation}, a joint bin--group estimator improves validation log loss after both main effects for 43.0\%/73.3\%/83.6\% of eligible AutoML-A/E/Criteo pairs. This diagnostic uses no \scalarlens output; the ablation suite tests whether the model exploits the residual structure, and the main matrix tests transfer. Appendix~\ref{app:context-evidence} gives the full audit.

\subsection{Ablation Study}
\label{sec:information-ablation}

All variants share data, backbone, seeds, focal basis, and output shape; source masking preserves nominal tensors. Table~\ref{tab:ablation-suite} reports changes from the complete method across AutoML-A/E/Criteo with DCNv2 and DeepFM (six cells, 18 paired seeds). Negative values mean that removing or replacing a pathway hurts. The blocks test information source, internal mechanism, and matched alternatives. ContextFiLM is an ablation, not an RQ1 peer.

\subsubsection{Interventions on Context Sources}
Removing response loses 0.0005--0.0091 AUC in all six cells. Adding Categorical context to Numerical context only improves all 18 paired seeds; other numerical fields (never the focal value) are auxiliary, improving 11/18 pairs and four means beyond Categorical context only.

Local Only removes both response paths. Variants that retain only one source preserve the same focal coordinate and isolate which family of surrounding fields supports recovery. The results identify categorical context as reliable and numerical context as auxiliary rather than uniformly necessary.

\subsubsection{Component Interventions}
NoNorm, StopGrad, and $T{=}1$ lose 0.0037, 0.0025, and 0.0011 AUC on average and are negative in every cell. FrozenMessage is mixed ($-0.0002$ average), making message recomputation an optimization choice rather than a defining claim.

Because RMS normalization follows coordinate construction, its gain cannot come from relocating the focal value. StopGrad and $T{=}1$ instead isolate joint learning and response depth, identifying both as functional rather than decorative choices.

\subsubsection{Matched Alternatives}
DEER+Norm and LocalMLP each lose 0.0043 on average across the six settings, rejecting stabilization and local capacity as explanations. ContextFiLM loses 0.0011 on average, with five losses and one narrow win; \scalarlens claims consistency, not dominance over every conditioner in every setting.

These controls respectively add stabilization, local nonlinear capacity, and affine conditioning with one update. Their deficits rule out three simpler explanations for the gain from separating coordinate and response.

Together, the blocks identify the coupled interface: the focal scalar owns its coordinate, categorical context supplies the reliable response signal, and normalization, joint adaptation of all components, and multiple updates make that response usable. Numerical context remains auxiliary on real CTR data. Local Only, source interventions, and matched alternatives respectively test coordinate sufficiency, response information, and simpler explanations.

\subsection{Controlled Mechanism Recovery and Context Shift}
\label{sec:controlled-recovery}

\begin{figure}[t]
  \centering
  \includegraphics[width=\columnwidth]{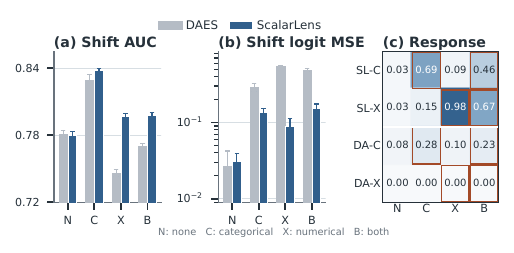}
  \caption{\textbf{Controlled recovery under context shift.} Bars compare DAES and ScalarLens in AUC on the shifted test set (a) and MSE relative to the true logit (b); whiskers denote the standard deviation over five seeds. Panel (c) reports token displacement after categorical or numerical intervention; orange outlines mark sources active in the generator. N/C/X/B denote no interaction, categorical interaction, numerical interaction, and both interactions. Both methods feed the same linear consumer, and the shift changes context prevalence but not $p(y\mid\mathbf{x},\mathbf{c})$.}
  \Description{Two grouped bar panels compare AUC on the shifted test set and error in recovering the true logit for DAES and ScalarLens under four controlled mechanisms. A heat map shows token displacement caused by categorical and numerical context. ScalarLens leads under categorical, numerical, and mixed mechanisms; DAES has zero response to interventions on numerical context by design.}
  \label{fig:mechanism-recovery}
\end{figure}

To separate response recovery from a powerful consumer, a controlled probe uses three numerical and two categorical fields, a known logit with no interaction or with categorical, numerical, or mixed interactions, and a shared linear consumer. A shifted test set reverses the prevalence and correlations of contexts while preserving $p(y\mid\mathbf{x},\mathbf{c})$. Experiments with five seeds compare five methods. Figure~\ref{fig:mechanism-recovery} shows DAES, the closest conditional peer, on the more difficult split. Appendix~\ref{app:synthetic-generator} specifies the complete data-generating process; the artifact provides the executable generator and all results.

Training, validation, and IID test data share one context distribution; the shifted split changes only covariates and retains scalar support. Hence differences between methods within a split reflect recovery of the same known response rather than a changed prediction rule. Exposing categorical tokens to the linear consumer also makes the control conservative: a numerical encoder invariant to context can use additive category effects, but cannot delegate an unmodeled interaction between numbers and context to a downstream MLP.

With no interaction, \scalarlens IID AUC ($0.7797\!\pm\!0.0035$) matches DEER/DAES. Under categorical, numerical, and mixed mechanisms, it reaches AUC values of 0.8383, 0.7969, and 0.7983 on the shifted test set, compared with 0.8296, 0.7463, and 0.7703 for DAES. It also reduces categorical logit MSE from 0.2891 to 0.1349. Numerical intervention moves its focal token by 0.9824 in the numerical mechanism and 0.6703 in the mixed mechanism, compared with zero for DAES, which uses only categorical context. The retained focal coordinate changes by exactly zero. Thus the response changes without relocating the coordinate.

The case without interactions bounds the explanation based on additional capacity. In the interacting cases, token displacement follows the active source while the coordinate audit remains zero, which aligns predictive recovery with the claimed mechanism.

\paragraph{Rarity stress test on real data.}
For a stress test on real data, training frequencies over Criteo's 26 categorical fields define row surprisal without using labels. Validation data determine the thresholds for common and rare groups, which are then applied unchanged to test covariates. Frozen checkpoints rank \scalarlens first in all four DNN/DCNv2 groups and every paired seed (Table~\ref{tab:context-rarity-main}); its margin in rare contexts (0.0012--0.0016) exceeds the margin in common contexts (0.0007). This experiment tests frequency rarity, not temporal shift.

Subgroup labels never influence training, model selection, or threshold fitting on test labels; the audit measures frozen representations rather than creating another tuning opportunity.

\begin{table}[t]
\caption{\textbf{Performance of frozen checkpoints in common and rare Criteo contexts.} The best baseline is selected by unrounded mean among DAES, DEER, and NaryDis. Values are test AUC mean$_{\pm\mathrm{std}}$ over three seeds; $\Delta$ is ScalarLens minus that baseline.}
\label{tab:context-rarity-main}
\centering
\singlecoltableformat
\begin{tabular*}{\columnwidth}{@{\extracolsep{\fill}}llcccc@{}}
\toprule
Backbone & Group & ScalarLens & Best baseline & $\Delta$ & Wins \\
\midrule
DNN & Common & \textbf{\aucstd{.8152}{.0001}} & \aucstd{.8145}{.0001} (NaryDis) & +.0007 & 3/3 \\
DNN & Rare & \textbf{\aucstd{.8160}{.0002}} & \aucstd{.8148}{.0000} (DEER) & +.0012 & 3/3 \\
DCNv2 & Common & \textbf{\aucstd{.8169}{.0000}} & \aucstd{.8162}{.0001} (DEER) & +.0007 & 3/3 \\
DCNv2 & Rare & \textbf{\aucstd{.8174}{.0003}} & \aucstd{.8158}{.0004} (DAES) & +.0016 & 3/3 \\
\bottomrule
\end{tabular*}
\end{table}

\subsection{Parameter Sensitivity}
\label{sec:sensitivity}

\begin{figure*}[t]
  \centering
  \includegraphics[width=\textwidth]{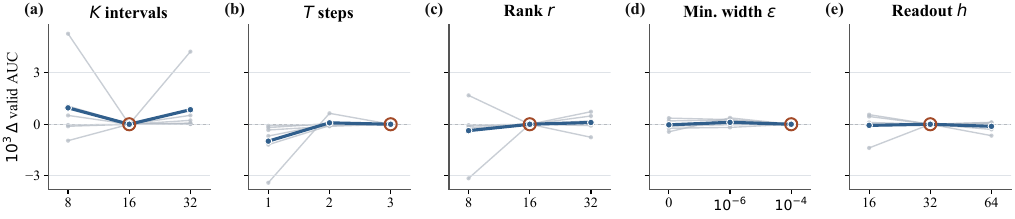}
  \caption{\textbf{Parameter sensitivity of \scalarlens.} The change in validation AUC from the frozen default is plotted for interval count $K$, response steps $T$, rank $r$ of the low-rank operator, minimum interval width, and readout width $h$. The vertical marker denotes the frozen default.}
  \Description{Five sensitivity panels show light curves for six combinations of datasets and backbones and a darker mean curve. Performance is broadly stable around the frozen defaults, while one response step is consistently worse.}
  \label{fig:sensitivity}
\end{figure*}

Across six combinations of datasets and backbones and three seeds, Figure~\ref{fig:sensitivity} shows broad plateaus for $K$, rank, minimum width, and readout width; $T=1$ is consistently weaker. We retain the interior frozen default $(K,T,r,h)=(16,3,16,32)$ rather than promote an endpoint from one favorable setting. These validation results never revise the rankings on test data.

Two or three response steps recover the deficit of $T=1$; other parameters occupy shallow plateaus. The apparent gain at $K=8$ is concentrated in one setting, and $K=32$ fails the declared promotion rule. The evidence therefore supports one interior default rather than tuning each dataset separately.

\subsection{Accuracy and Efficiency}
\label{sec:efficiency}

We compare matched training and inference cost over the same six settings; baselines share one implementation regime rather than individually tuned deployment limits.

\begin{figure}[t]
  \centering
  \includegraphics[width=\columnwidth]{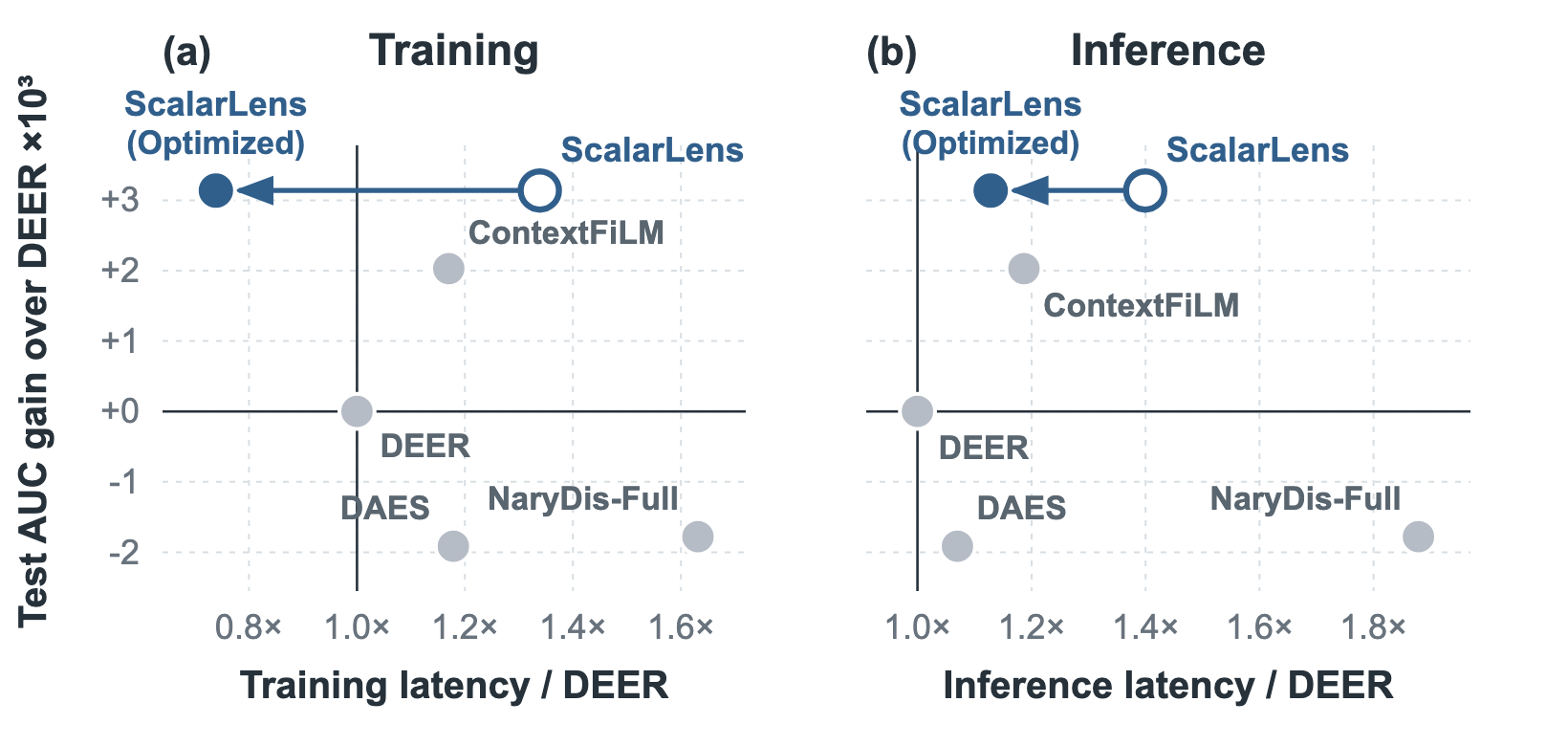}
  \caption{\textbf{Accuracy and efficiency during training and inference.} Panels (a) and (b) show latency for one training step and for inference with a batch of 4096, respectively. Every method is averaged over AutoML-A/E/Criteo with DCNv2 and DeepFM; both axes are relative to DEER. \emph{\scalarlens} uses the common eager execution path. \emph{\scalarlens{} (Optimized)} applies an optimization that preserves the function and is measured separately in each of the same six settings. Its unchanged vertical coordinate indicates that execution optimization preserves accuracy.}
  \Description{Two scatter plots compare gain in test AUC with relative training and inference latency. ScalarLens and its optimized implementation share the highest accuracy coordinate; optimization moves the point to the left of DEER during training and close to DEER during inference.}
  \label{fig:runtime-boundary}
\end{figure}

Optimization moves \scalarlens from $1.34\times$ to $0.74\times$ DEER training latency and from $1.40\times$ to $1.13\times$ inference latency without changing AUC. Packed batches, prefetching, batched lookups, compilation of the complete training step, and fused Adam remove most training overhead; the remaining 13\% inference premium reflects the additional computation required for contextual responses.

The eager and optimized points separate representation cost from execution overhead. The optimization preserves the function, reduces training latency below that of DEER, and approaches but does not surpass DEER during inference.

\section{Discussion and Limitations}

\scalarlens provides a transferable representation interface, but not a universally dominant predictor. Categorical context is reliable, whereas numerical context is auxiliary on real data but decisive in the controlled mechanism. The AutoInt, ContextFiLM, and FrozenMessage exceptions identify settings in which simpler designs remain competitive.

Evaluation on original numerical scales is primary because production features are often generated, versioned, and refreshed independently. Avoiding transformed copies reduces storage and I/O, while avoiding an online affine stage simplifies synchronization between training and serving. This advantage neither makes normalization arithmetic intrinsically expensive nor removes schema validation. The standardized rerun controls for input scale.

\paragraph{Deployment scope.}
The systems argument concerns lifecycle cost, not the storage required by two moments. Offline materialization adds derived columns and backfills; online transformation adds a versioned operator to training and serving. \scalarlens instead stores fitted ranges and the learned mesh with the checkpoint. It still performs interpolation and contextual computation, whose latency is measured in Figure~\ref{fig:runtime-boundary}. The advantage is consolidation of scale handling in an auditable model artifact.

AutoML-A/E share one family, all datasets informed development, and the 27 settings are correlated. Random splits and the controlled shift do not establish temporal, causal, or transfer to an unseen family. Such claims require temporal splits, delayed feedback, and new domains. Our supported conclusion is architectural: a stable coordinate with a bounded contextual response works across the tested consumers.

\paragraph{Evidence hierarchy.}
Matched ablations, controlled recovery, frozen rarity groups, and the scale audit support the 25/27 count. Their agreement is stronger than any single statistic. FrozenMessage and the auxiliary numerical path identify deployment simplifications; neither changes the reported matrix.

\section{Conclusion}

\scalarlens separates a learned coordinate determined only by the scalar from a bounded contextual response, while changing only numerical embeddings. It accepts original values and ranks first in 25 of 27 combinations of datasets and backbones. Ablations rule out scale correction and local capacity as sufficient explanations, while controlled mechanisms recover categorical, numerical, and mixed responses without moving the coordinate. Numerical representations should therefore specify both an invariant object and an adaptive object rather than require one token to serve both roles. This separation provides a reusable design principle and a concrete invariant for auditing future CTR encoders.

\clearpage
\enlargethispage{3pt}
\bibliographystyle{ACM-Reference-Format}
\bibliography{references}

\clearpage
\appendix
\twocolumn[
\begin{center}
  {\LARGE\bfseries Supplementary Material for ScalarLens}
\end{center}
\vspace{0.5em}
]
\section{Controlled Data-Generating Process}
\label{app:synthetic-generator}

Each example contains $x=(x_0,x_1,x_2)\in[-1,1]^3$ and categories $c_0\in\{0,1,2,3\}$ and $c_1\in\{0,1,2\}$. Let
\begin{align}
q(x,c)&=0.65\sin(\pi x_0)+0.35x_1-0.25x_2\nonumber\\
&\quad+0.18(c_0-1.5)-0.12(c_1-1.0).
\end{align}
Labels follow $y\sim\operatorname{Bernoulli}(\sigma(q+r_s-0.18))$, where the four response mechanisms are
\begingroup\small
\begin{align}
r_{\mathrm{add}}&=1.10\sin(1.35\pi x_0),\nonumber\\
r_{\mathrm{cat}}&=(0.85+0.22c_0+0.14c_1)
 \sin(\pi[x_0+0.16(c_0-1.5)\nonumber\\
&\hspace{8.5em}-0.11(c_1-1)]),\nonumber\\
r_{\mathrm{num}}&=1.25\sin\!\bigl(\pi[x_0+0.58x_1]\bigr)+1.05x_0x_2,\nonumber\\
r_{\mathrm{mix}}&=(0.72+0.17c_0+0.11c_1+0.28\tanh(2x_2))\nonumber\\
&\quad\cdot\sin(\pi[x_0+0.13(c_0-1.5)-0.09(c_1-1)\nonumber\\
&\hspace{8.5em}+0.45x_1])+0.85x_0x_2.
\label{eq:synthetic-generator}
\end{align}
\endgroup
For training, validation, and IID test data, category probabilities are $(0.52,0.27,0.15,0.06)$ and $(0.62,0.28,0.10)$; the shifted test reverses both vectors. Starting from $\widetilde x\sim\mathcal U([-1,1]^3)$, we set $x_0=\widetilde x_0$, $x_1=\operatorname{clip}(0.78\widetilde x_1+d\,0.16(c_0-1.5))$, and $x_2=\operatorname{clip}(0.82\widetilde x_2+d\,0.14(c_1-1))$, with $d=1$ for training, validation, and IID test and $d=-1$ for the shifted test. Thus only covariate prevalence and correlation change; Equation~\eqref{eq:synthetic-generator} is fixed. Split sizes are 20k/5k/10k/10k, and all methods use identical draws for each of five seeds.

\section{Evidence for Context Sources}
\label{app:context-evidence}

The diagnostic extends Figure~\ref{fig:motivation} to every eligible pair. A numerical target is eligible when its training-quantile discretization contains at least three distinct nonempty bins and at least one candidate context also forms at least three groups. Constant and low-cardinality numerical fields that collapse below this threshold are excluded before candidate scoring. This rule retains 13 of 23 numerical targets on AutoML-A, 6 of 6 on AutoML-E, and 11 of 13 on Criteo. Training quantiles define focal bins $b$ and context groups $c$. An empirical Bayes additive estimator models both marginal effects, while a smoothed joint table permits residual dependence:
\begin{align}
\operatorname{logit}p_{\mathrm{add}}(y{=}1\mid b,c)
&=\operatorname{logit}p(y{=}1\mid b)+\operatorname{logit}p(y{=}1\mid c)\nonumber\\
&\quad-\operatorname{logit}p(y{=}1),\nonumber\\
p_{\mathrm{joint}}(b,c)
&=\frac{y_{bc}+200p_{\mathrm{add}}(b,c)}{n_{bc}+200}.
\label{eq:appendix-joint}
\end{align}
A deterministic sample of 600,000 training rows is split to fit candidates and select one context for each target; the selected pair is fitted again on the full sample and evaluated once on validation data. Selection never uses validation labels, and the test set remains unopened. Table~\ref{tab:context-crossfit} reports the resulting sign tests across target fields and bootstrap intervals.

\begin{table}[t]
\caption{Context suitability selected on training data and confirmed on validation data. Positive $\Delta$LL means that the joint table of target and context lowers validation log loss relative to an additive model containing both marginal effects. Confidence intervals bootstrap target fields; $p$ is from a two sided exact sign test. Test data are not used.}
\label{tab:context-crossfit}
\centering
\singlecoltablesize
\setlength{\tabcolsep}{3.2pt}
\renewcommand{\arraystretch}{1.08}
\resizebox{\columnwidth}{!}{%
\begin{tabular}{llrrrr}
\toprule
Context source & Dataset & Positive targets & Mean $\Delta$LL & 95\% bootstrap CI & Sign $p$ \\
\midrule
Numerical & AutoML-A & 12/13 & 0.000725 & [0.000479, 0.000965] & 0.00342 \\
Numerical & AutoML-E & 6/6 & 0.000618 & [0.000540, 0.000694] & 0.0312 \\
Numerical & Criteo & 11/11 & 0.010581 & [0.007156, 0.013884] & 0.000977 \\
Numerical & Pooled & 29/30 & 0.004317 & [0.002340, 0.006559] & $5.77\!\times\!10^{-8}$ \\
\midrule
Categorical & AutoML-A & 10/13 & 0.000308 & [0.000078, 0.000618] & 0.0923 \\
Categorical & AutoML-E & 6/6 & 0.000502 & [0.000374, 0.000624] & 0.0312 \\
Categorical & Criteo & 11/11 & 0.002573 & [0.001270, 0.004050] & 0.000977 \\
Categorical & Pooled & 27/30 & 0.001177 & [0.000594, 0.001885] & $8.43\!\times\!10^{-6}$ \\
\bottomrule
\end{tabular}
}
\end{table}

The joint estimator improves validation log loss for 43.0\% of 663 pairs on AutoML-A, 73.3\% of 150 on AutoML-E, and 83.6\% of 286 on Criteo; selected categorical and numerical contexts are positive for 27/30 and 29/30 targets. This establishes conditional structure, not equal neural utility: Table~\ref{tab:ablation-suite} identifies categorical context as reliable and other numerical fields as auxiliary.

\subsection{Matched neural checks and interventions on frozen checkpoints}

The $2\times2$ intervention on validation data in Table~\ref{tab:context-neural-validation} adds each context source under both states of the other source. Categorical context is positive in every paired run; other numerical fields have a smaller effect that depends on the setting.

\begin{table}[t]
\caption{Matched neural $2\times2$ study on validation AUC. Effects add the named source under the stated condition and pool AutoML-A/E, DCNv2/DeepFM, and three seeds. The interval bootstraps the 12 paired runs; $p$ is from a one sided exact sign test.}
\label{tab:context-neural-validation}
\centering
\singlecoltableformat
\begin{tabular*}{\columnwidth}{@{\extracolsep{\fill}}p{.15\columnwidth}p{.14\columnwidth}P{.17\columnwidth}P{.12\columnwidth}P{.12\columnwidth}P{.19\columnwidth}@{}}
\toprule
Added source & Existing source & $10^3\!\times$ mean $\Delta$AUC & Positive pairs & Sign $p$ & 95\% bootstrap CI \\
\midrule
Numerical & None & +0.066 & 10/12 & 0.0193 & [$-0.367$, $+0.409$] \\
Numerical & Categorical & +0.151 & 6/12 & 0.6128 & [$-0.128$, $+0.453$] \\
Categorical & None & +5.779 & 12/12 & 0.000244 & [$+4.457$, $+7.330$] \\
Categorical & Numerical & +5.865 & 12/12 & 0.000244 & [$+4.389$, $+7.600$] \\
\bottomrule
\end{tabular*}
\end{table}

Masking during training may change the learned optimum, so we also freeze each Full checkpoint and permute numerical context within bins of the focal value. This preserves the focal coordinate while breaking the alignment between rows. Table~\ref{tab:context-checkpoint-intervention} shows that the reinjected embedding changes in all settings, while predictive reliance is concentrated on Criteo.

\begin{table}[t]
\caption{Intervention on numerical context at a fixed checkpoint, evaluated on validation data. A positive AUC drop or increase in log loss means that permutation within bins of the target value harms the frozen model. Relative RMS measures the change in the reinjected target embedding; the final column counts seeds with increased log loss.}
\label{tab:context-checkpoint-intervention}
\centering
\singlecoltableformat
\begin{tabular*}{\columnwidth}{@{\extracolsep{\fill}}p{.13\columnwidth}p{.12\columnwidth}P{.12\columnwidth}P{.13\columnwidth}P{.13\columnwidth}P{.16\columnwidth}P{.08\columnwidth}@{}}
\toprule
Dataset & Backbone & $10^3\!\times$ AUC drop & $10^3\!\times$ LL increase & Prediction $|\Delta|$ & Embedding rel. RMS & LL worse \\
\midrule
AutoML-A & DCNv2 & $-0.0079$ & $+0.0004$ & 0.000082 & 0.3188\% & 1/3 \\
AutoML-A & DeepFM & $+0.0095$ & $+0.0035$ & 0.000117 & 0.5007\% & 2/3 \\
AutoML-E & DCNv2 & $-0.0104$ & $-0.0008$ & 0.000055 & 0.2336\% & 2/3 \\
AutoML-E & DeepFM & $-0.0167$ & $+0.0003$ & 0.000067 & 0.3658\% & 2/3 \\
Criteo & DCNv2 & $+0.4217$ & $+0.4623$ & 0.005193 & 11.0441\% & 3/3 \\
Criteo & DeepFM & $+0.2532$ & $+0.2252$ & 0.004303 & 10.2173\% & 3/3 \\
\bottomrule
\end{tabular*}
\end{table}

\section{Detailed Context and Robustness Audits}
\label{app:factorial-effects}
\label{app:robustness-audits}

With $A_{nc}$ denoting AUC under numerical source $n$ and categorical source $c$, categorical context is positive in all 18 paired runs, whereas numerical context is smaller and depends on the setting. Permuting context at a frozen checkpoint changes the reinjected numerical token throughout, but predictive reliance is concentrated on Criteo. The artifact contains all conditional increments, interactions, and interventions for each seed. This distinction prevents a dependency observed in data from being overstated as uniform utility for the neural model.

For categorical field $j$, let $n_j(v)$ be the training count of ID $v$ and $V_j$ its vocabulary size. We measure the rarity of a Criteo context by row surprisal computed without labels:
\begin{equation}
R(\mathbf c)=-\frac{1}{26}\sum_{j=1}^{26}\log\frac{n_j(c_j)+1}{N+V_j}.
\end{equation}
Validation covariates determine thresholds at the 20th and 80th percentiles. Applying these thresholds unchanged to the test set yields 916,010 common and 918,284 rare examples. Table~\ref{tab:context-rarity-main} gives the compact comparison in the main paper; Table~\ref{tab:context-rarity} reports every frozen method.

\begin{table}[t]
\caption{\textbf{Stress test of frozen checkpoints under rare Criteo contexts.} Frequencies come from training data; validation covariates determine thresholds for common and rare groups. Values are test AUC mean$_{\pm\mathrm{std}}$ over three seeds. $\Delta_{\mathrm F}$ is AUC in the rare group relative to Field.}
\label{tab:context-rarity}
\centering
\singlecoltableformat
\begin{tabular*}{\columnwidth}{@{\extracolsep{\fill}}llccc@{}}
\toprule
Backbone & Method & Common & Rare & $\Delta_{\mathrm F}$ \\
\midrule
\multirow{5}{*}{DNN}
 & Field         & \aucstd{.8110}{.0004} & \aucstd{.8074}{.0016} & --- \\
 & DAES          & \aucstd{.8141}{.0001} & \aucstd{.8146}{.0002} & +.0072 \\
 & DEER          & \aucstd{.8145}{.0001} & \aucstd{.8148}{.0000} & +.0073 \\
 & NaryDis  & \aucstd{.8145}{.0001} & \aucstd{.8146}{.0002} & +.0071 \\
 & \textbf{\scalarlens} & \textbf{\aucstd{.8152}{.0001}} & \textbf{\aucstd{.8160}{.0002}} & \textbf{+.0085} \\
\midrule
\multirow{5}{*}{DCNv2}
 & Field         & \aucstd{.8086}{.0006} & \aucstd{.8063}{.0011} & --- \\
 & DAES          & \aucstd{.8159}{.0000} & \aucstd{.8158}{.0004} & +.0095 \\
 & DEER          & \aucstd{.8162}{.0001} & \aucstd{.8157}{.0001} & +.0094 \\
 & NaryDis  & \aucstd{.8161}{.0001} & \aucstd{.8157}{.0002} & +.0094 \\
 & \textbf{\scalarlens} & \textbf{\aucstd{.8169}{.0000}} & \textbf{\aucstd{.8174}{.0003}} & \textbf{+.0111} \\
\bottomrule
\end{tabular*}
\end{table}

\section{Reproducibility and Scope}
\label{app:method-details}

\paragraph{Frozen protocol and run accounting.}
Each full matrix contains $3\times9\times19\times3=1{,}539$ runs. All methods share row partitions, numerical inputs, categorical IDs, optimizer, seeds, early stopping, and the policy of evaluating the test set only after model selection; failed jobs are retried under the same configuration. Aggregation rejects missing or empty records and ranks a method only after all three seeds are present.

\paragraph{Model boundary and leakage controls.}
Every method replaces only the numerical representation interface. Original categorical lookup embeddings remain unchanged. DeepFM, WideDeep, and NFM expose the same 16-dimensional numerical vector to their linear and interaction branches, preventing an unencoded scalar from bypassing the method. Vocabularies, ranges, quantile boundaries, and diagnostics are fitted without test labels; test metrics are computed from checkpoints selected on validation data.

\paragraph{Artifact and execution environment.}
The artifact contains the module, AutoML preprocessing entry point, dataset manifests, configuration snapshots and digests, records for each seed, aggregation and plotting scripts, raw latency traces, and equivalence tests for eager and optimized execution. Efficiency uses one PPU-ZW810E accelerator, PyTorch~2.6.0, Python~3.10, and FP32 without TF32. Each timing averages three independent processes. Because the frozen harness did not record peak allocated memory, we make no claim about memory efficiency.

\paragraph{Policy for sensitivity analyses and negative results.}
The sensitivity study changes one parameter specific to \scalarlens at a time while fixing learning rate, batch size, optimizer, backbone widths, early stopping, data order, and seeds. No tested candidate satisfied the declared rule requiring broad improvement strongly enough to replace the frozen default. We retain AutoInt settings in which \scalarlens does not win, mixed interventions on numerical context, and weak FrozenMessage results. These outcomes delimit the claim: stable coordinates with contextual responses provide a reliable interface in the tested matrix, not a guarantee that every context source, response mechanism, or consumer benefits.

\subsection{Benchmark on original numerical scales and secondary standardization audit}

Under the original numerical ranges, \scalarlens ranks first in 25 of 27 combinations of datasets and backbones and second in the remaining two, with mean rank 1.074. It wins eight of nine backbones on AutoML-A, eight of nine on AutoML-E, and all nine on Criteo. Its mean AUC on each dataset exceeds the strongest competitor by 0.0070, 0.0037, and 0.0012, respectively, while its mean Logloss is also lowest on all three datasets.

As a secondary control, the shared z-score rerun fitted only on training data removes all 60 cells below 0.55 AUC in the original scale evaluation. After stabilization, \scalarlens still wins 18/27 paired cells against DEER, 21/27 against DAES, and 25/27 against NaryDis. Mean paired gains are $+0.0031$, $+0.0047$, and $+0.0103$ AUC, with all three comparisons significant after Holm correction. It wins 16/27 standardized cells and lies in the top three for 19/27. This audit shows that the primary result is not reducible to compatibility with numerical ranges; it is not the primary deployment protocol.

\clearpage
\onecolumn

\begin{center}
\begin{minipage}{\textwidth}
\captionof{table}{Secondary audit of input scale: complete test results under one shared z-score protocol fitted only on training data. The layout and notation match Table~\ref{tab:full-results}: means use four decimals, subscripts in muted blue are standard deviations over three seeds, and bold and underline mark the two highest untruncated AUCs in each column. Global Rank and Wins summarize all 27 combinations of datasets and backbones and are shown only in the final dataset block.}
\label{tab:zscore-full-results}
\centering
\tiny
\setlength{\tabcolsep}{0.8pt}
\renewcommand{\arraystretch}{0.80}
\resizebox{\textwidth}{!}{%
\begin{tabular}{lcccccccccccccccccccccc}
\toprule
\multirow{2}{*}{Method} & \multicolumn{2}{c}{DNN} & \multicolumn{2}{c}{DeepFM} & \multicolumn{2}{c}{WideDeep} & \multicolumn{2}{c}{NFM} & \multicolumn{2}{c}{PNN} & \multicolumn{2}{c}{DCN} & \multicolumn{2}{c}{DCNv2} & \multicolumn{2}{c}{AutoInt} & \multicolumn{2}{c}{FiBiNET} & \multicolumn{2}{c}{Mean} & \multicolumn{2}{c}{Global} \\
\cmidrule(lr){2-3} \cmidrule(lr){4-5} \cmidrule(lr){6-7} \cmidrule(lr){8-9} \cmidrule(lr){10-11} \cmidrule(lr){12-13} \cmidrule(lr){14-15} \cmidrule(lr){16-17} \cmidrule(lr){18-19} \cmidrule(lr){20-21} \cmidrule(lr){22-23}
 & AUC & LL & AUC & LL & AUC & LL & AUC & LL & AUC & LL & AUC & LL & AUC & LL & AUC & LL & AUC & LL & AUC & LL & Rank & Wins \\
\midrule
\multicolumn{23}{l}{\textbf{AutoML-A}} \\
Field & \aucstd{0.7033}{.0034} & 0.2406 & \aucstd{0.7074}{.0020} & 0.2391 & \aucstd{0.7062}{.0012} & 0.2384 & \aucstd{0.6851}{.0017} & 0.2286 & \aucstd{0.7008}{.0030} & 0.2405 & \aucstd{0.7077}{.0022} & 0.2409 & \aucstd{0.7083}{.0017} & 0.2433 & \aucstd{0.7112}{.0058} & 0.2104 & \aucstd{0.6972}{.0027} & 0.2443 & \aucstd{0.7030}{.0009} & 0.2362 & -- & -- \\
Linear & \aucstd{0.7052}{.0014} & 0.2447 & \aucstd{0.6965}{.0009} & 0.2319 & \aucstd{0.7035}{.0048} & 0.2436 & \aucstd{0.6886}{.0020} & 0.2221 & \aucstd{0.7062}{.0044} & 0.2475 & \aucstd{0.7042}{.0022} & 0.2387 & \aucstd{0.7055}{.0013} & 0.2356 & \aucstd{0.7076}{.0007} & 0.2311 & \aucstd{\underline{0.7099}}{.0027} & 0.2387 & \aucstd{0.7030}{.0006} & 0.2371 & -- & -- \\
Linear-ReLU & \aucstd{0.7037}{.0002} & 0.2413 & \aucstd{0.5675}{.1169} & 22.7809 & \aucstd{0.7022}{.0022} & 0.2404 & \aucstd{0.6825}{.0073} & 0.2263 & \aucstd{0.7049}{.0022} & 0.2470 & \aucstd{0.7032}{.0013} & 0.2383 & \aucstd{0.7039}{.0012} & 0.2380 & \aucstd{0.7081}{.0023} & 0.2340 & \aucstd{0.7082}{.0019} & 0.2458 & \aucstd{0.6871}{.0125} & 2.7436 & -- & -- \\
Quantile bucket & \aucstd{0.7077}{.0008} & 0.2407 & \aucstd{0.7080}{.0025} & 0.2381 & \aucstd{0.7064}{.0024} & 0.2403 & \aucstd{0.6903}{.0030} & 0.2253 & \aucstd{0.7066}{.0011} & 0.2387 & \aucstd{0.7046}{.0004} & 0.2448 & \aucstd{0.7055}{.0012} & 0.2460 & \aucstd{0.7096}{.0049} & 0.2192 & \aucstd{0.7073}{.0027} & 0.2528 & \aucstd{0.7051}{.0011} & 0.2384 & -- & -- \\
Log-squared bucket & \aucstd{0.6799}{.0019} & 0.2424 & \aucstd{0.6817}{.0014} & 0.2490 & \aucstd{0.6795}{.0017} & 0.2527 & \aucstd{0.6637}{.0029} & 0.2350 & \aucstd{0.6775}{.0062} & 0.2529 & \aucstd{0.6785}{.0009} & 0.2535 & \aucstd{0.6817}{.0028} & 0.2487 & \aucstd{0.6845}{.0023} & 0.2240 & \aucstd{0.6791}{.0026} & 0.2486 & \aucstd{0.6785}{.0003} & 0.2452 & -- & -- \\
AutoDis & \aucstd{0.7061}{.0007} & 0.2400 & \aucstd{0.7066}{.0047} & 0.2364 & \aucstd{0.7057}{.0039} & 0.2379 & \aucstd{0.6912}{.0012} & 0.2299 & \aucstd{0.7060}{.0023} & 0.2374 & \aucstd{0.7035}{.0028} & 0.2475 & \aucstd{0.7081}{.0030} & 0.2462 & \aucstd{\underline{0.7137}}{.0008} & 0.2132 & \aucstd{0.7060}{.0044} & 0.2500 & \aucstd{0.7052}{.0011} & 0.2376 & -- & -- \\
PLE-A & \aucstd{0.7082}{.0005} & 0.2422 & \aucstd{0.6746}{.0087} & 0.2348 & \aucstd{0.7011}{.0044} & 0.2468 & \aucstd{0.6888}{.0047} & 0.2164 & \aucstd{0.7103}{.0003} & 0.2349 & \aucstd{0.7063}{.0017} & 0.2393 & \aucstd{0.7049}{.0013} & 0.2307 & \aucstd{0.7077}{.0038} & 0.2415 & \aucstd{0.7047}{.0030} & 0.2318 & \aucstd{0.7007}{.0018} & 0.2354 & -- & -- \\
PLE-B & \aucstd{0.7072}{.0017} & 0.2412 & \aucstd{0.7031}{.0029} & 0.2191 & \aucstd{0.7047}{.0027} & 0.2343 & \aucstd{0.6946}{.0035} & 0.2230 & \aucstd{\underline{0.7105}}{.0017} & 0.2381 & \aucstd{0.7025}{.0018} & 0.2564 & \aucstd{\underline{0.7098}}{.0017} & 0.2353 & \aucstd{0.7081}{.0014} & 0.2470 & \aucstd{0.7071}{.0036} & 0.2348 & \aucstd{0.7053}{.0014} & 0.2366 & -- & -- \\
PLE-ReLU & \aucstd{0.7043}{.0014} & 0.2411 & \aucstd{0.6961}{.0006} & 0.2195 & \aucstd{0.7038}{.0016} & 0.2357 & \aucstd{0.6880}{.0029} & 0.2262 & \aucstd{0.7066}{.0023} & 0.2374 & \aucstd{0.6980}{.0031} & 0.2551 & \aucstd{0.7071}{.0025} & 0.2354 & \aucstd{0.7044}{.0044} & 0.2500 & \aucstd{0.7073}{.0038} & 0.2390 & \aucstd{0.7017}{.0010} & 0.2377 & -- & -- \\
Periodic & \aucstd{0.7032}{.0015} & 0.2550 & \aucstd{0.7011}{.0028} & 0.2246 & \aucstd{0.7025}{.0009} & 0.2516 & \aucstd{\underline{0.6954}}{.0072} & 0.2202 & \aucstd{0.7033}{.0021} & 0.2460 & \aucstd{0.7048}{.0021} & 0.2553 & \aucstd{0.7092}{.0026} & 0.2331 & \aucstd{0.7088}{.0006} & 0.2423 & \aucstd{0.7072}{.0021} & 0.2354 & \aucstd{0.7039}{.0015} & 0.2404 & -- & -- \\
PLR & \aucstd{0.7079}{.0024} & 0.2457 & \aucstd{0.7078}{.0009} & 0.2381 & \aucstd{0.7067}{.0020} & 0.2437 & \aucstd{0.6862}{.0048} & 0.2329 & \aucstd{0.7047}{.0053} & 0.2489 & \aucstd{0.7043}{.0034} & 0.2481 & \aucstd{\textbf{0.7100}}{.0016} & 0.2367 & \aucstd{0.7125}{.0034} & 0.2225 & \aucstd{0.7055}{.0011} & 0.2376 & \aucstd{0.7051}{.0015} & 0.2394 & -- & -- \\
B-spline & \aucstd{0.7076}{.0025} & 0.2467 & \aucstd{\underline{0.7101}}{.0016} & 0.2333 & \aucstd{\underline{0.7096}}{.0020} & 0.2347 & \aucstd{0.6897}{.0020} & 0.2234 & \aucstd{0.7066}{.0024} & 0.2370 & \aucstd{\underline{0.7078}}{.0014} & 0.2394 & \aucstd{0.7086}{.0019} & 0.2389 & \aucstd{0.7111}{.0007} & 0.2214 & \aucstd{0.7021}{.0026} & 0.2527 & \aucstd{0.7059}{.0003} & 0.2364 & -- & -- \\
NaryDis-Lite & \aucstd{0.6919}{.0021} & 0.2420 & \aucstd{0.6892}{.0029} & 0.2446 & \aucstd{0.6894}{.0023} & 0.2436 & \aucstd{0.6761}{.0028} & 0.2260 & \aucstd{0.6911}{.0030} & 0.2550 & \aucstd{0.6868}{.0041} & 0.2523 & \aucstd{0.6888}{.0016} & 0.2451 & \aucstd{0.7062}{.0041} & 0.2188 & \aucstd{0.6874}{.0018} & 0.2461 & \aucstd{0.6897}{.0005} & 0.2415 & -- & -- \\
NaryDis & \aucstd{0.6889}{.0012} & 0.2488 & \aucstd{0.6892}{.0040} & 0.2412 & \aucstd{0.6896}{.0045} & 0.2397 & \aucstd{0.6753}{.0029} & 0.2310 & \aucstd{0.6893}{.0017} & 0.2469 & \aucstd{0.6907}{.0042} & 0.2494 & \aucstd{0.6874}{.0032} & 0.2498 & \aucstd{0.7018}{.0060} & 0.2161 & \aucstd{0.6882}{.0047} & 0.2599 & \aucstd{0.6889}{.0021} & 0.2425 & -- & -- \\
DEER & \aucstd{\textbf{0.7090}}{.0008} & 0.2414 & \aucstd{0.7095}{.0020} & 0.2403 & \aucstd{0.7089}{.0009} & 0.2426 & \aucstd{0.6920}{.0031} & 0.2259 & \aucstd{0.7079}{.0014} & 0.2422 & \aucstd{0.7070}{.0011} & 0.2396 & \aucstd{0.7072}{.0022} & 0.2403 & \aucstd{\textbf{0.7153}}{.0019} & 0.2222 & \aucstd{0.7064}{.0014} & 0.2486 & \aucstd{\underline{0.7070}}{.0005} & 0.2381 & -- & -- \\
DAE & \aucstd{0.7061}{.0018} & 0.2340 & \aucstd{0.7053}{.0016} & 0.2371 & \aucstd{0.7053}{.0013} & 0.2371 & \aucstd{0.6904}{.0022} & 0.2283 & \aucstd{0.7036}{.0014} & 0.2426 & \aucstd{0.7038}{.0022} & 0.2446 & \aucstd{0.7043}{.0020} & 0.2395 & \aucstd{0.7107}{.0029} & 0.2199 & \aucstd{0.6998}{.0045} & 0.2426 & \aucstd{0.7033}{.0003} & 0.2362 & -- & -- \\
DAES-Gate & \aucstd{0.7057}{.0014} & 0.2418 & \aucstd{0.7036}{.0043} & 0.2402 & \aucstd{0.7062}{.0036} & 0.2387 & \aucstd{0.6883}{.0091} & 0.2228 & \aucstd{0.7038}{.0027} & 0.2498 & \aucstd{0.7039}{.0022} & 0.2417 & \aucstd{0.7064}{.0034} & 0.2386 & \aucstd{0.7109}{.0019} & 0.2251 & \aucstd{0.7070}{.0058} & 0.2344 & \aucstd{0.7040}{.0003} & 0.2370 & -- & -- \\
DAES-Tran & \aucstd{0.7033}{.0027} & 0.2496 & \aucstd{0.7007}{.0056} & 0.2376 & \aucstd{0.7014}{.0040} & 0.2401 & \aucstd{0.6845}{.0067} & 0.2288 & \aucstd{0.7050}{.0040} & 0.2469 & \aucstd{0.7040}{.0012} & 0.2446 & \aucstd{0.7012}{.0036} & 0.2412 & \aucstd{0.7073}{.0043} & 0.2196 & \aucstd{0.7047}{.0039} & 0.2410 & \aucstd{0.7013}{.0005} & 0.2388 & -- & -- \\
\textbf{ScalarLens} & \aucstd{\underline{0.7082}}{.0022} & 0.2371 & \aucstd{\textbf{0.7151}}{.0016} & 0.2169 & \aucstd{\textbf{0.7135}}{.0011} & 0.2186 & \aucstd{\textbf{0.7165}}{.0009} & 0.2182 & \aucstd{\textbf{0.7139}}{.0003} & 0.2334 & \aucstd{\textbf{0.7138}}{.0010} & 0.2405 & \aucstd{0.7088}{.0033} & 0.2275 & \aucstd{0.7123}{.0027} & 0.2331 & \aucstd{\textbf{0.7154}}{.0030} & 0.2173 & \aucstd{\textbf{0.7131}}{.0005} & 0.2270 & -- & -- \\
\midrule
\multicolumn{23}{l}{\textbf{AutoML-E}} \\
Field & \aucstd{0.8356}{.0033} & 0.0718 & \aucstd{0.8394}{.0008} & 0.0706 & \aucstd{\underline{0.8395}}{.0007} & 0.0705 & \aucstd{0.8398}{.0007} & 0.0706 & \aucstd{0.8389}{.0014} & 0.0699 & \aucstd{0.8351}{.0038} & 0.0706 & \aucstd{0.8375}{.0019} & 0.0696 & \aucstd{\textbf{0.8444}}{.0011} & 0.0682 & \aucstd{0.8344}{.0010} & 0.0731 & \aucstd{0.8383}{.0004} & 0.0706 & -- & -- \\
Linear & \aucstd{0.8383}{.0005} & 0.0708 & \aucstd{0.8377}{.0023} & 0.0705 & \aucstd{0.8367}{.0015} & 0.0713 & \aucstd{0.8378}{.0007} & 0.0712 & \aucstd{0.8379}{.0032} & 0.0725 & \aucstd{0.8372}{.0013} & 0.0719 & \aucstd{0.8369}{.0008} & 0.0721 & \aucstd{0.8393}{.0012} & 0.0704 & \aucstd{0.8339}{.0050} & 0.0719 & \aucstd{0.8373}{.0012} & 0.0714 & -- & -- \\
Linear-ReLU & \aucstd{0.8383}{.0007} & 0.0705 & \aucstd{0.8369}{.0008} & 0.0707 & \aucstd{0.8366}{.0008} & 0.0712 & \aucstd{0.8360}{.0022} & 0.0714 & \aucstd{0.8385}{.0025} & 0.0719 & \aucstd{0.8370}{.0010} & 0.0716 & \aucstd{0.8367}{.0013} & 0.0718 & \aucstd{0.8407}{.0009} & 0.0697 & \aucstd{0.8336}{.0066} & 0.0721 & \aucstd{0.8371}{.0012} & 0.0712 & -- & -- \\
Quantile bucket & \aucstd{\underline{0.8396}}{.0002} & 0.0703 & \aucstd{0.8387}{.0023} & 0.0708 & \aucstd{0.8374}{.0022} & 0.0708 & \aucstd{\underline{0.8402}}{.0010} & 0.0705 & \aucstd{\underline{0.8400}}{.0009} & 0.0708 & \aucstd{0.8342}{.0034} & 0.0724 & \aucstd{0.8364}{.0018} & 0.0709 & \aucstd{0.8429}{.0005} & 0.0690 & \aucstd{0.8364}{.0011} & 0.0712 & \aucstd{0.8384}{.0004} & 0.0708 & -- & -- \\
Log-squared bucket & \aucstd{0.8356}{.0033} & 0.0705 & \aucstd{0.8371}{.0010} & 0.0701 & \aucstd{0.8375}{.0011} & 0.0702 & \aucstd{0.8384}{.0004} & 0.0703 & \aucstd{0.8374}{.0010} & 0.0709 & \aucstd{0.8335}{.0051} & 0.0718 & \aucstd{0.8353}{.0028} & 0.0704 & \aucstd{0.8409}{.0009} & 0.0681 & \aucstd{0.8314}{.0061} & 0.0741 & \aucstd{0.8363}{.0016} & 0.0707 & -- & -- \\
AutoDis & \aucstd{0.8371}{.0019} & 0.0712 & \aucstd{0.8385}{.0010} & 0.0698 & \aucstd{0.8383}{.0008} & 0.0701 & \aucstd{0.8396}{.0011} & 0.0705 & \aucstd{0.8382}{.0021} & 0.0702 & \aucstd{0.8379}{.0011} & 0.0708 & \aucstd{0.8380}{.0007} & 0.0704 & \aucstd{0.8431}{.0012} & 0.0690 & \aucstd{0.8338}{.0022} & 0.0728 & \aucstd{0.8383}{.0004} & 0.0705 & -- & -- \\
PLE-A & \aucstd{0.8383}{.0014} & 0.0714 & \aucstd{0.8389}{.0005} & 0.0708 & \aucstd{0.8380}{.0018} & 0.0715 & \aucstd{0.8374}{.0012} & 0.0705 & \aucstd{0.8394}{.0017} & 0.0716 & \aucstd{0.8381}{.0019} & 0.0713 & \aucstd{\underline{0.8394}}{.0008} & 0.0718 & \aucstd{0.8402}{.0003} & 0.0702 & \aucstd{0.8362}{.0008} & 0.0722 & \aucstd{0.8384}{.0005} & 0.0713 & -- & -- \\
PLE-B & \aucstd{0.8365}{.0019} & 0.0713 & \aucstd{0.8365}{.0035} & 0.0719 & \aucstd{0.8369}{.0014} & 0.0713 & \aucstd{0.8376}{.0004} & 0.0711 & \aucstd{0.8395}{.0002} & 0.0720 & \aucstd{0.8375}{.0026} & 0.0710 & \aucstd{0.8382}{.0006} & 0.0713 & \aucstd{0.8407}{.0014} & 0.0702 & \aucstd{0.8339}{.0018} & 0.0736 & \aucstd{0.8375}{.0009} & 0.0715 & -- & -- \\
PLE-ReLU & \aucstd{0.8370}{.0013} & 0.0706 & \aucstd{0.8374}{.0023} & 0.0710 & \aucstd{0.8360}{.0014} & 0.0710 & \aucstd{0.8381}{.0004} & 0.0710 & \aucstd{0.8384}{.0007} & 0.0714 & \aucstd{0.8372}{.0027} & 0.0706 & \aucstd{0.8380}{.0009} & 0.0725 & \aucstd{0.8398}{.0013} & 0.0714 & \aucstd{0.8365}{.0008} & 0.0725 & \aucstd{0.8376}{.0006} & 0.0713 & -- & -- \\
Periodic & \aucstd{0.8345}{.0005} & 0.0732 & \aucstd{\underline{0.8395}}{.0005} & 0.0692 & \aucstd{0.8381}{.0019} & 0.0704 & \aucstd{0.8392}{.0004} & 0.0706 & \aucstd{0.8370}{.0020} & 0.0719 & \aucstd{\underline{0.8382}}{.0016} & 0.0703 & \aucstd{0.8383}{.0008} & 0.0713 & \aucstd{0.8390}{.0015} & 0.0704 & \aucstd{0.8342}{.0032} & 0.0727 & \aucstd{0.8375}{.0001} & 0.0711 & -- & -- \\
PLR & \aucstd{0.8384}{.0023} & 0.0704 & \aucstd{0.8390}{.0006} & 0.0696 & \aucstd{0.8372}{.0004} & 0.0702 & \aucstd{0.8379}{.0013} & 0.0703 & \aucstd{0.8383}{.0009} & 0.0704 & \aucstd{0.8377}{.0011} & 0.0715 & \aucstd{0.8374}{.0002} & 0.0706 & \aucstd{0.8396}{.0015} & 0.0685 & \aucstd{0.8305}{.0040} & 0.0746 & \aucstd{0.8373}{.0008} & 0.0707 & -- & -- \\
B-spline & \aucstd{0.8387}{.0017} & 0.0703 & \aucstd{0.8392}{.0014} & 0.0703 & \aucstd{0.8391}{.0014} & 0.0702 & \aucstd{0.8389}{.0003} & 0.0699 & \aucstd{0.8363}{.0021} & 0.0713 & \aucstd{0.8379}{.0021} & 0.0701 & \aucstd{0.8355}{.0013} & 0.0700 & \aucstd{\underline{0.8437}}{.0005} & 0.0698 & \aucstd{0.8300}{.0055} & 0.0744 & \aucstd{0.8377}{.0008} & 0.0707 & -- & -- \\
NaryDis-Lite & \aucstd{0.8376}{.0006} & 0.0710 & \aucstd{0.8365}{.0012} & 0.0711 & \aucstd{0.8366}{.0013} & 0.0712 & \aucstd{0.8382}{.0002} & 0.0712 & \aucstd{0.8369}{.0006} & 0.0714 & \aucstd{0.8360}{.0021} & 0.0721 & \aucstd{0.8345}{.0003} & 0.0726 & \aucstd{0.8407}{.0020} & 0.0687 & \aucstd{0.8329}{.0015} & 0.0720 & \aucstd{0.8366}{.0005} & 0.0713 & -- & -- \\
NaryDis & \aucstd{0.8349}{.0016} & 0.0723 & \aucstd{0.8372}{.0013} & 0.0703 & \aucstd{0.8365}{.0017} & 0.0704 & \aucstd{0.8390}{.0005} & 0.0713 & \aucstd{0.8363}{.0025} & 0.0706 & \aucstd{0.8346}{.0035} & 0.0708 & \aucstd{0.8364}{.0012} & 0.0705 & \aucstd{0.8423}{.0008} & 0.0693 & \aucstd{0.8330}{.0035} & 0.0733 & \aucstd{0.8367}{.0006} & 0.0710 & -- & -- \\
DEER & \aucstd{0.8377}{.0009} & 0.0710 & \aucstd{0.8387}{.0016} & 0.0701 & \aucstd{0.8388}{.0016} & 0.0698 & \aucstd{0.8398}{.0003} & 0.0707 & \aucstd{0.8382}{.0016} & 0.0707 & \aucstd{0.8375}{.0034} & 0.0716 & \aucstd{0.8377}{.0015} & 0.0707 & \aucstd{0.8435}{.0008} & 0.0682 & \aucstd{\underline{0.8365}}{.0019} & 0.0721 & \aucstd{\underline{0.8387}}{.0003} & 0.0705 & -- & -- \\
DAE & \aucstd{0.8370}{.0014} & 0.0699 & \aucstd{0.8380}{.0006} & 0.0702 & \aucstd{0.8380}{.0018} & 0.0700 & \aucstd{0.8391}{.0002} & 0.0705 & \aucstd{0.8391}{.0015} & 0.0700 & \aucstd{0.8373}{.0019} & 0.0707 & \aucstd{0.8377}{.0007} & 0.0703 & \aucstd{0.8429}{.0009} & 0.0686 & \aucstd{0.8321}{.0010} & 0.0744 & \aucstd{0.8379}{.0004} & 0.0705 & -- & -- \\
DAES-Gate & \aucstd{0.8387}{.0009} & 0.0704 & \aucstd{0.8356}{.0023} & 0.0720 & \aucstd{0.8341}{.0037} & 0.0720 & \aucstd{0.8389}{.0014} & 0.0711 & \aucstd{0.8396}{.0015} & 0.0709 & \aucstd{0.8375}{.0023} & 0.0705 & \aucstd{0.8373}{.0014} & 0.0709 & \aucstd{0.8423}{.0013} & 0.0689 & \aucstd{0.8301}{.0015} & 0.0741 & \aucstd{0.8371}{.0001} & 0.0712 & -- & -- \\
DAES-Tran & \aucstd{0.8372}{.0018} & 0.0715 & \aucstd{0.8358}{.0062} & 0.0718 & \aucstd{0.8362}{.0052} & 0.0713 & \aucstd{0.8384}{.0008} & 0.0703 & \aucstd{0.8350}{.0015} & 0.0729 & \aucstd{0.8364}{.0022} & 0.0708 & \aucstd{0.8373}{.0022} & 0.0711 & \aucstd{0.8424}{.0004} & 0.0694 & \aucstd{0.8296}{.0050} & 0.0722 & \aucstd{0.8365}{.0012} & 0.0713 & -- & -- \\
\textbf{ScalarLens} & \aucstd{\textbf{0.8416}}{.0009} & 0.0690 & \aucstd{\textbf{0.8429}}{.0013} & 0.0680 & \aucstd{\textbf{0.8426}}{.0005} & 0.0679 & \aucstd{\textbf{0.8419}}{.0004} & 0.0683 & \aucstd{\textbf{0.8425}}{.0010} & 0.0694 & \aucstd{\textbf{0.8409}}{.0009} & 0.0687 & \aucstd{\textbf{0.8426}}{.0007} & 0.0700 & \aucstd{0.8417}{.0008} & 0.0695 & \aucstd{\textbf{0.8415}}{.0006} & 0.0682 & \aucstd{\textbf{0.8420}}{.0002} & 0.0688 & -- & -- \\
\midrule
\multicolumn{23}{l}{\textbf{Criteo}} \\
Field & \aucstd{0.8086}{.0001} & 0.4433 & \aucstd{0.8056}{.0001} & 0.4461 & \aucstd{0.8079}{.0001} & 0.4441 & \aucstd{0.8008}{.0003} & 0.4508 & \aucstd{0.8092}{.0000} & 0.4426 & \aucstd{0.8082}{.0003} & 0.4437 & \aucstd{0.8094}{.0001} & 0.4434 & \aucstd{0.8090}{.0001} & 0.4429 & \aucstd{0.8090}{.0001} & 0.4430 & \aucstd{0.8075}{.0000} & 0.4444 & 10.815 & 1/27 \\
Linear & \aucstd{0.8086}{.0001} & 0.4432 & \aucstd{0.8041}{.0002} & 0.4479 & \aucstd{0.8077}{.0002} & 0.4444 & \aucstd{0.8009}{.0003} & 0.4506 & \aucstd{0.8090}{.0001} & 0.4429 & \aucstd{0.8083}{.0000} & 0.4438 & \aucstd{0.8094}{.0001} & 0.4427 & \aucstd{0.8087}{.0000} & 0.4433 & \aucstd{0.8095}{.0002} & 0.4428 & \aucstd{0.8074}{.0001} & 0.4446 & 12.778 & 0/27 \\
Linear-ReLU & \aucstd{0.8093}{.0001} & 0.4427 & \aucstd{0.8053}{.0002} & 0.4468 & \aucstd{0.8085}{.0001} & 0.4438 & \aucstd{0.8022}{.0002} & 0.4493 & \aucstd{0.8096}{.0001} & 0.4428 & \aucstd{0.8092}{.0001} & 0.4428 & \aucstd{0.8108}{.0001} & 0.4414 & \aucstd{0.8092}{.0001} & 0.4427 & \aucstd{0.8102}{.0002} & 0.4421 & \aucstd{0.8082}{.0000} & 0.4438 & 12.889 & 0/27 \\
Quantile bucket & \aucstd{0.8102}{.0001} & 0.4418 & \aucstd{0.8076}{.0001} & 0.4441 & \aucstd{0.8094}{.0002} & 0.4428 & \aucstd{0.8052}{.0002} & 0.4467 & \aucstd{0.8113}{.0001} & 0.4405 & \aucstd{0.8102}{.0001} & 0.4418 & \aucstd{0.8115}{.0001} & 0.4405 & \aucstd{0.8105}{.0002} & 0.4414 & \aucstd{0.8108}{.0001} & 0.4413 & \aucstd{0.8096}{.0000} & 0.4423 & 6.778 & 0/27 \\
Log-squared bucket & \aucstd{0.7936}{.0002} & 0.4567 & \aucstd{0.7913}{.0002} & 0.4586 & \aucstd{0.7929}{.0001} & 0.4574 & \aucstd{0.7899}{.0002} & 0.4600 & \aucstd{0.7945}{.0000} & 0.4556 & \aucstd{0.7938}{.0001} & 0.4563 & \aucstd{0.7947}{.0000} & 0.4558 & \aucstd{0.7939}{.0002} & 0.4563 & \aucstd{0.7938}{.0001} & 0.4565 & \aucstd{0.7931}{.0001} & 0.4570 & 17.370 & 0/27 \\
AutoDis & \aucstd{0.8102}{.0001} & 0.4418 & \aucstd{0.8077}{.0004} & 0.4440 & \aucstd{0.8095}{.0001} & 0.4426 & \aucstd{0.8043}{.0002} & 0.4475 & \aucstd{0.8109}{.0001} & 0.4412 & \aucstd{0.8103}{.0000} & 0.4417 & \aucstd{0.8116}{.0001} & 0.4406 & \aucstd{0.8105}{.0001} & 0.4413 & \aucstd{0.8105}{.0001} & 0.4419 & \aucstd{0.8095}{.0001} & 0.4425 & 7.222 & 0/27 \\
PLE-A & \aucstd{0.8105}{.0001} & 0.4415 & \aucstd{0.8066}{.0004} & 0.4451 & \aucstd{0.8097}{.0001} & 0.4423 & \aucstd{0.8050}{.0004} & 0.4469 & \aucstd{0.8114}{.0000} & 0.4406 & \aucstd{0.8104}{.0001} & 0.4419 & \aucstd{0.8122}{.0001} & 0.4399 & \aucstd{0.8105}{.0001} & 0.4414 & \aucstd{\underline{0.8117}}{.0003} & 0.4402 & \aucstd{0.8098}{.0000} & 0.4422 & 7.741 & 0/27 \\
PLE-B & \aucstd{0.8105}{.0003} & 0.4415 & \aucstd{0.8065}{.0003} & 0.4456 & \aucstd{\underline{0.8100}}{.0002} & 0.4420 & \aucstd{0.8053}{.0002} & 0.4466 & \aucstd{\underline{0.8117}}{.0001} & 0.4402 & \aucstd{\textbf{0.8107}}{.0001} & 0.4414 & \aucstd{\underline{0.8125}}{.0001} & 0.4396 & \aucstd{0.8107}{.0001} & 0.4413 & \aucstd{\textbf{0.8117}}{.0000} & 0.4403 & \aucstd{0.8100}{.0001} & 0.4421 & 7.148 & 2/27 \\
PLE-ReLU & \aucstd{0.8084}{.0001} & 0.4435 & \aucstd{0.8039}{.0001} & 0.4487 & \aucstd{0.8079}{.0000} & 0.4440 & \aucstd{0.8008}{.0000} & 0.4507 & \aucstd{0.8091}{.0001} & 0.4429 & \aucstd{0.8083}{.0001} & 0.4439 & \aucstd{0.8097}{.0000} & 0.4427 & \aucstd{0.8088}{.0000} & 0.4433 & \aucstd{0.8094}{.0001} & 0.4428 & \aucstd{0.8074}{.0000} & 0.4447 & 12.926 & 0/27 \\
Periodic & \aucstd{0.8093}{.0003} & 0.4429 & \aucstd{0.8040}{.0002} & 0.4498 & \aucstd{0.8084}{.0001} & 0.4436 & \aucstd{0.8031}{.0004} & 0.4488 & \aucstd{0.8101}{.0002} & 0.4419 & \aucstd{0.8092}{.0002} & 0.4426 & \aucstd{0.8109}{.0002} & 0.4411 & \aucstd{0.8091}{.0004} & 0.4427 & \aucstd{0.8100}{.0001} & 0.4423 & \aucstd{0.8082}{.0001} & 0.4440 & 10.407 & 0/27 \\
PLR & \aucstd{0.8097}{.0001} & 0.4420 & \aucstd{0.8064}{.0002} & 0.4454 & \aucstd{0.8090}{.0001} & 0.4433 & \aucstd{0.8037}{.0000} & 0.4479 & \aucstd{0.8103}{.0002} & 0.4414 & \aucstd{0.8095}{.0002} & 0.4422 & \aucstd{0.8113}{.0000} & 0.4410 & \aucstd{0.8098}{.0002} & 0.4421 & \aucstd{0.8103}{.0001} & 0.4419 & \aucstd{0.8089}{.0001} & 0.4430 & 9.222 & 1/27 \\
B-spline & \aucstd{0.8078}{.0003} & 0.4441 & \aucstd{0.8053}{.0003} & 0.4462 & \aucstd{0.8075}{.0003} & 0.4444 & \aucstd{0.7984}{.0001} & 0.4530 & \aucstd{0.8084}{.0004} & 0.4439 & \aucstd{0.8080}{.0002} & 0.4439 & \aucstd{0.8091}{.0002} & 0.4432 & \aucstd{0.8082}{.0004} & 0.4435 & \aucstd{0.8091}{.0004} & 0.4431 & \aucstd{0.8069}{.0001} & 0.4450 & 10.593 & 0/27 \\
NaryDis-Lite & \aucstd{0.8093}{.0001} & 0.4425 & \aucstd{0.8065}{.0002} & 0.4451 & \aucstd{0.8087}{.0001} & 0.4435 & \aucstd{0.8018}{.0001} & 0.4496 & \aucstd{0.8099}{.0000} & 0.4419 & \aucstd{0.8091}{.0001} & 0.4428 & \aucstd{0.8110}{.0001} & 0.4411 & \aucstd{0.8098}{.0001} & 0.4421 & \aucstd{0.8099}{.0001} & 0.4422 & \aucstd{0.8084}{.0001} & 0.4434 & 14.593 & 0/27 \\
NaryDis & \aucstd{0.8098}{.0001} & 0.4420 & \aucstd{0.8069}{.0001} & 0.4447 & \aucstd{0.8091}{.0002} & 0.4431 & \aucstd{0.8023}{.0002} & 0.4493 & \aucstd{0.8102}{.0001} & 0.4416 & \aucstd{0.8097}{.0002} & 0.4425 & \aucstd{0.8111}{.0002} & 0.4409 & \aucstd{0.8097}{.0001} & 0.4422 & \aucstd{0.8102}{.0000} & 0.4418 & \aucstd{0.8088}{.0000} & 0.4431 & 14.000 & 0/27 \\
DEER & \aucstd{\textbf{0.8108}}{.0001} & 0.4413 & \aucstd{\textbf{0.8082}}{.0002} & 0.4437 & \aucstd{\textbf{0.8101}}{.0002} & 0.4421 & \aucstd{0.8052}{.0002} & 0.4466 & \aucstd{\textbf{0.8118}}{.0001} & 0.4401 & \aucstd{\underline{0.8106}}{.0001} & 0.4417 & \aucstd{0.8122}{.0002} & 0.4398 & \aucstd{\textbf{0.8111}}{.0000} & 0.4409 & \aucstd{0.8111}{.0001} & 0.4409 & \aucstd{\underline{0.8101}}{.0000} & 0.4419 & 4.222 & 7/27 \\
DAE & \aucstd{0.8099}{.0000} & 0.4420 & \aucstd{0.8076}{.0003} & 0.4451 & \aucstd{0.8094}{.0002} & 0.4428 & \aucstd{0.8049}{.0002} & 0.4471 & \aucstd{0.8106}{.0002} & 0.4420 & \aucstd{0.8100}{.0003} & 0.4420 & \aucstd{0.8112}{.0002} & 0.4412 & \aucstd{0.8102}{.0002} & 0.4416 & \aucstd{0.8103}{.0001} & 0.4419 & \aucstd{0.8093}{.0000} & 0.4429 & 9.296 & 0/27 \\
DAES-Gate & \aucstd{\underline{0.8105}}{.0002} & 0.4413 & \aucstd{0.8078}{.0001} & 0.4440 & \aucstd{0.8099}{.0002} & 0.4422 & \aucstd{\underline{0.8057}}{.0001} & 0.4464 & \aucstd{0.8117}{.0001} & 0.4402 & \aucstd{0.8105}{.0002} & 0.4415 & \aucstd{0.8123}{.0001} & 0.4398 & \aucstd{\underline{0.8107}}{.0001} & 0.4415 & \aucstd{0.8113}{.0001} & 0.4407 & \aucstd{0.8101}{.0001} & 0.4420 & 7.778 & 0/27 \\
DAES-Tran & \aucstd{0.8102}{.0001} & 0.4416 & \aucstd{0.8071}{.0000} & 0.4449 & \aucstd{0.8095}{.0003} & 0.4426 & \aucstd{0.8055}{.0001} & 0.4462 & \aucstd{0.8107}{.0001} & 0.4414 & \aucstd{0.8102}{.0001} & 0.4418 & \aucstd{0.8118}{.0002} & 0.4405 & \aucstd{0.8102}{.0001} & 0.4421 & \aucstd{0.8109}{.0000} & 0.4414 & \aucstd{0.8096}{.0001} & 0.4425 & 11.370 & 0/27 \\
\textbf{ScalarLens} & \aucstd{0.8103}{.0002} & 0.4418 & \aucstd{\underline{0.8078}}{.0001} & 0.4439 & \aucstd{0.8099}{.0001} & 0.4423 & \aucstd{\textbf{0.8089}}{.0001} & 0.4430 & \aucstd{0.8116}{.0001} & 0.4405 & \aucstd{0.8079}{.0002} & 0.4440 & \aucstd{\textbf{0.8133}}{.0001} & 0.4387 & \aucstd{0.8107}{.0001} & 0.4411 & \aucstd{0.8114}{.0001} & 0.4407 & \aucstd{\textbf{0.8102}}{.0000} & 0.4418 & \textbf{2.815} & \textbf{16/27} \\
\bottomrule
\end{tabular}%
}
\end{minipage}
\end{center}

\vspace{0.3em}
\noindent
\begin{minipage}[t]{0.485\textwidth}
\paragraph{Controlled response recovery.}
With a shared linear consumer and an unchanged conditional label rule, \scalarlens reaches AUC values of 0.8383, 0.7969, and 0.7983 on the shifted test set for categorical, numerical, and mixed response mechanisms, compared with 0.8296, 0.7463, and 0.7703 for DAES. In the categorical mechanism, MSE relative to the true logit falls from 0.2891 to 0.1349. Numerical intervention moves the learned response token by 0.9824 in the numerical generator and 0.6703 in the mixed generator, while the retained focal coordinate changes by exactly zero. Predictive recovery and internal intervention therefore support the same separation of coordinate and response.
\end{minipage}\hfill
\begin{minipage}[t]{0.485\textwidth}
\paragraph{Evidence from rare contexts and efficiency.}
Across all six pairs of backbones and seeds in the frozen Criteo rare group, \scalarlens exceeds DAES, DEER, and NaryDis. Mean paired gains are $+0.0015$, $+0.0014$, and $+0.0015$ AUC, with all three paired bootstrap intervals excluding zero; its margin over the strongest baseline is larger for rare than common contexts. The optimized implementation preserves predictions, loss, gradients, and test AUC while reducing training latency from $1.34\times$ to $0.74\times$ DEER and inference latency from $1.40\times$ to $1.13\times$. Thus the strongest accuracy point remains practical after systems optimization.
\end{minipage}

\vspace{0.45em}
\paragraph{Coherent ablation evidence.}
The positive results are supported by matched removals rather than the main matrix alone. Relative to the full model, Local Only loses in all six combinations of datasets and backbones by an average of 0.0038 AUC; removing normalization, stopping contextual gradients, and reducing the response to one step lose 0.0037, 0.0025, and 0.0011 on average. Alternatives matched for capacity and preprocessing also fail to reproduce the complete interface: DEER+Norm and LocalMLP each lose about 0.0043 AUC on average, while ContextFiLM loses 0.0011 and wins only one of six settings. Together with the controlled coordinate audit, this pattern attributes the gain to the coupling between stable coordinates and contextual responses rather than to scale correction, additional local parameters, or an arbitrary conditioner.

\end{document}